\documentclass[useAMS,usenatbib]{mn2e}

\usepackage{graphicx}
\usepackage{amsmath}
\usepackage{multirow}
\usepackage{tabularx}
\usepackage{lscape}

\title[Planets and Stellar Activity: Hide and Seek in the CoRoT-7 system]{Planets and Stellar Activity: Hide and Seek in the CoRoT-7 system\thanks{Based on observations made with the HARPS instrument on the 3.6 m telescope under the program ID 088.C-0323 at Cerro La Silla (Chile).}}

\author[R. D. Haywood]{R. D. Haywood$^{1}$\thanks{E-mail:
rdh4@st-andrews.ac.uk}, A.Collier Cameron$^{1}$, D. Queloz$^{2}$, S.C.C. Barros$^{3}$, M. Deleuil$^{3}$, 
\newauthor
R. Fares$^{1}$,  M. Gillon$^{4}$, A.F. Lanza$^{5}$, C. Lovis$^{2}$, C. Moutou$^{3}$, F. Pepe$^{2}$,  
\newauthor
D. Pollacco$^{6}$, A. Santerne$^{3, 7}$, D. S\'{e}gransan$^{2}$ and Y. C. Unruh$^{8}$ \\
$^{1}$SUPA, School of Physics and Astronomy, University of St Andrews, St Andrews KY16 9SS, UK \\
$^{2}$Observatoire de Gen\`{e}ve, 51 Ch. des Maillettes, 1290 Sauverny, Switzerland \\
$^{3}$Aix Marseille Universit\'e, CNRS, LAM (Laboratoire d'Astrophysique de Marseille) UMR 7326, 13388, Marseille, France \\
$^{4}$Institut d'Astrophysique et de G\'{e}ophysique, Universit\'{e} de Li\`{e}ge, All\'{e}e du 6 ao\^{u}t 17, Bat. B5C, 4000 Li\`{e}ge, Belgium\\
$^{5}$INAF-Osservatorio Astrofisico di Catania, via S. Sofia, 78 - 95123 Catania. Italy\\
$^{6}$Department of Physics, University of Warwick, Coventry CV4 7AL, UK\\
$^{7}$Centro de Astrof\'isica, Universidade do Porto, Rua das Estrelas, 4150-762 Porto, Portugal\\
$^{8}$Astrophysics Group, Blackett Laboratory, Imperial College London, London SW7 2AZ, UK}

\begin{document}

\date{Accepted 2014 July 2. Received 2014 June 30; in original form 2013 May 30}

\pagerange{\pageref{firstpage}--\pageref{lastpage}} \pubyear{2002}

\maketitle

\label{firstpage}

\begin{abstract}

Since the discovery of the transiting super-Earth CoRoT-7b, several investigations have yielded different results for the number and masses of planets present in the system, mainly owing to the star's high level of activity.
We re-observed CoRoT-7 in January 2012 with both HARPS and CoRoT, so that we now have the benefit of simultaneous radial-velocity and photometric data. This allows us to use the off-transit variations in the star's light curve to estimate the radial-velocity variations induced by the suppression of convective blueshift and the flux blocked by starspots. To account for activity-related effects in the radial-velocities which do not have a photometric signature, we also include an additional activity term in the radial-velocity model, which we treat as a Gaussian process with the same covariance properties (and hence the same frequency structure) as the light curve.
Our model was incorporated into a Monte Carlo Markov Chain in order to make a precise determination of the orbits of CoRoT-7b and CoRoT-7c. We measure the masses of planets b and c to be $4.73 \pm 0.95 \, $M$_{\oplus}$ and $13.56 \pm 1.08\, $M$_{\oplus}$, respectively. The density of CoRoT-7b is $(6.61 \pm 1.72)(R_p/1.58$ R$_{\oplus})^{-3}$ g.cm$^{-3}$, which is compatible with a rocky composition.
We search for evidence of an additional planet d, identified by previous authors with a period close to 9 days. 
We are not able to confirm the existence of a planet with this orbital period, which is close to the second harmonic of the stellar rotation at $\sim 7.9$ days. Using Bayesian model selection we find that  a model with two planets plus activity-induced variations is most favoured.

\end{abstract}


\section{Introduction}\label{intro}

In July 2009, \citet{Leger:2009cb} announced the discovery of the transiting planet CoRoT-7b with an orbital period of 0.85 days. At the time, it had the smallest exoplanetary radius ever measured, of $1.68\pm0.09 R_{\oplus}$.

Following this discovery, a 4-month intensive HARPS campaign was launched in order to measure the mass of CoRoT-7b. The results of this run are reported in \citet{Queloz:2009bo}. They expected the radial-velocity (hereafter RV) variations to be heavily affected by stellar activity, given the large modulations in the CoRoT photometry. The star's light curve (2008-2009 CoRoT run) shows modulations due to starspots of up to 2\%, which tells us that CoRoT-7 is more active than the Sun, whose greatest variations in irradiance recorded are of 0.34\% \citep{Kopp:2011dv}.
Indeed, a few simultaneous photometric measurements from the Euler Swiss telescope confirmed that CoRoT-7 was very spotted throughout the HARPS run. In order to remove the activity-induced RV variations from the data, \citet{Queloz:2009bo} applied a pre-whitening procedure followed by a harmonic decomposition. For the prewhitening, the period of the stellar rotation signal is identified by means of a Fourier analysis, and a sine fit with this period is subtracted from the data. This operation is applied to the residuals to remove the next strongest signal, and so on until the noise level is reached. All the signals detected with this method were determined to be associated with harmonics of the stellar rotation period, except two signals at 0.85 and 3.69 days. The RV signal at 0.85 days was found to be consistent with the CoRoT transit ephemeris, thus confirming the planetary nature of CoRoT-7b. Its mass was found to be $4.8\pm0.8$ M$_{\oplus}$.
In order to assess the nature of the signal at 3.69 days, \citet{Queloz:2009bo} used a harmonic decomposition to create a high pass filter: the RV data were fitted with a Fourier series comprising the first three harmonics of the stellar rotation period, within a time window sliding along the data. The length of this window (coherence time) was chosen to be 20 days, so that any signals varying over a longer timescale are filtered out -- starspots typically have lifetimes of about a month \citep{Schrijver:2002ht,Hussain:2002eg}. The harmonically filtered data were found to contain a strong periodic signal at 3.69 days, which was attributed to the orbit of CoRoT-7c, another super-Earth with a mass of $8.4\pm0.9$ M$_{\oplus}$. 

A few months later, \citet{Bruntt:2010iq} re-measured the stellar radius with improved stellar analysis techniques, which led to a slightly smaller planetary radius for CoRoT-7b than initially found, of $1.58\pm0.10 R_{\oplus}$.

A separate investigation was later carried out by \citet{Lanza:2010gx}. The stellar induced RV variations were synthesized based on a fit to the CoRoT light curve, which was computed according to a maximum entropy spot model \citep{Lanza:2009gd,Lanza:2011bfb}. The existence of the two planets was then confirmed by demonstrating that the activity-induced RV variations did not contain any spurious signals at the orbital periods of the two planets, with an estimated false alarm probability of less than $10^{-4}$.

In another analysis, \citet{Hatzes:2010jq} applied a prewhitening procedure to the full width at half-maximum (FWHM), bisector span and Ca II H\&K line emission derived from the HARPS spectra and cross-correlation analyses. These quantities vary according to activity only, and are independent of planetary orbital motions. No significant signals were found in any of these indicators at the periods of 0.85 and 3.69 days. Furthermore, they investigated the nature of a signal found in the RV data at 9.02 days. It had been previously detected by \citet{Queloz:2009bo} but had been attributed to a ``two frequency beating mode" resulting from an amplitude modulation of a signal at a period of 61 days. This is close to twice the stellar rotation period so it was deemed to be activity related.
Hatzes et al. found no trace of a signal at 9.02 days in any of the activity indicators. 
They thus suggest this RV signal could be attributed to a third planetary companion with a mass of $16.7\pm0.42$ M$_{\oplus}$. They also confirm the presence of CoRoT-7b and CoRoT-7c, but find different masses than calculated by \citet{Queloz:2009bo}. This is inevitable since the derived masses of planets are intimately connected with the methods used to mitigate the effects of stellar activity on the RV data. 

 \citet{Hatzes:2010jq,  Hatzes:2011fxa} developed a very simple method to remove stellar activity-induced RV variations, to obtain a more accurate mass for CoRoT-7b.
The method relies on making several well-separated observations on each night, which was the case for about half of the HARPS data. Under the assumption that the variations due to activity and other planets are negligible during the span of the observations on each night, it is possible to fit a keplerian orbit assuming that the velocity zero-point differs from night to night but remains constant within each night. 
\citet{Hatzes:2010jq} report a mass of CoRoT-7b of $6.9\pm1.4$ M$_{\oplus}$ and the second analysis  \citep{Hatzes:2011fxa} yields a mass of $7.42\pm 1.21$ M$_{\oplus}$, which is consistent.

Pont, Aigrain and Zucker (2010) carried out an analysis based on a maximum entropy spot model (similar to \citet{Lanza:2010gx}) which makes use of many small spots as opposed to few large spots. The model is constrained using FWHM and bisector information. A careful examination of the residuals of the activity and planet models led to the authors to add an additional noise term in order to account for possible systematics beyond the formal RV uncertainties. \citet{Pont:2010io} argue that CoRoT-7b is detected in the RV data with much less confidence than in previous analyses, and report a mass of $2.3 \pm 1.8$ M$_{\oplus}$ detected at a $1.2 \sigma$ level. Furthermore, they argued that the RV data are not numerous enough and lack the quality required to look for convincing evidence of additional companions.

\citet{Boisse:2011bw} applied their {\sc soap} tool (Boisse, Bonfils and Santos 2012) to the CoRoT-7 system. This program simulates spots on the surface of a rotating star and then uses this model to compute the activity-induced RV variations of the star. With this technique, they obtain mass estimates for CoRoT-7b and CoRoT-7c. They judge that their errors are underestimated and suggest adding a noise term of 1.5  ms$^{-1}$ to account for activity-driven RV variations. Their mass estimate for CoRoT-7b is in agreement with the value reported by \citet{Queloz:2009bo} but they find a slightly higher value for the mass of CoRoT-7c.

\citet{FerrazMello:2011gt} constructed their own version of the high-pass filter employed by \citet{Queloz:2009bo} in order to test the validity of this method and estimate masses for CoRoT-7b and 7c. They compared it to the method used by \citet{Hatzes:2010jq,Hatzes:2011fxa} and to a pure Fourier analysis. They concluded the method is robust, and obtained revised masses of $8.0\pm1.2$ M$_{\oplus}$ for CoRoT-7b and $13.6\pm1.4$ M$_{\oplus}$ for CoRoT-7c, but make no mention of CoRoT-7d.

The analysis by \citet{Lanza:2010gx}, which makes use of the CoRoT light curve \citep{Leger:2009cb} to model the activity-induced RV variations, and those by \citet{Pont:2010io} and \citet{Boisse:2011bw}, which rely on the tight correlation between the FWHM and the simultaneous Euler photometry \citep{Queloz:2009bo}, could be much improved with simultaneous photometric and RV data (see Lanza et al., in prep.). The spot activity on CoRoT-7 changes very rapidly and it is therefore not possible to deduce the form of the activity-driven RV variations from photometry taken up to a year before the RV data.
In the next section, we introduce the new simultaneous photometric and RV observations obtained in 2012 January with the CoRoT satellite and HARPS spectrograph.
In Section~\ref{totalmodel}, we describe our RV model which takes activity-induced RV variations into account by combining the method of Aigrain, Pont and Zucker (2012) and an additional RV Gaussian process.
Our model is implemented in Section~\ref{analysis}, and we discuss the outcomes in Section~\ref{discuss}.


\begin{figure*}
\centering
\includegraphics[width = \textwidth]{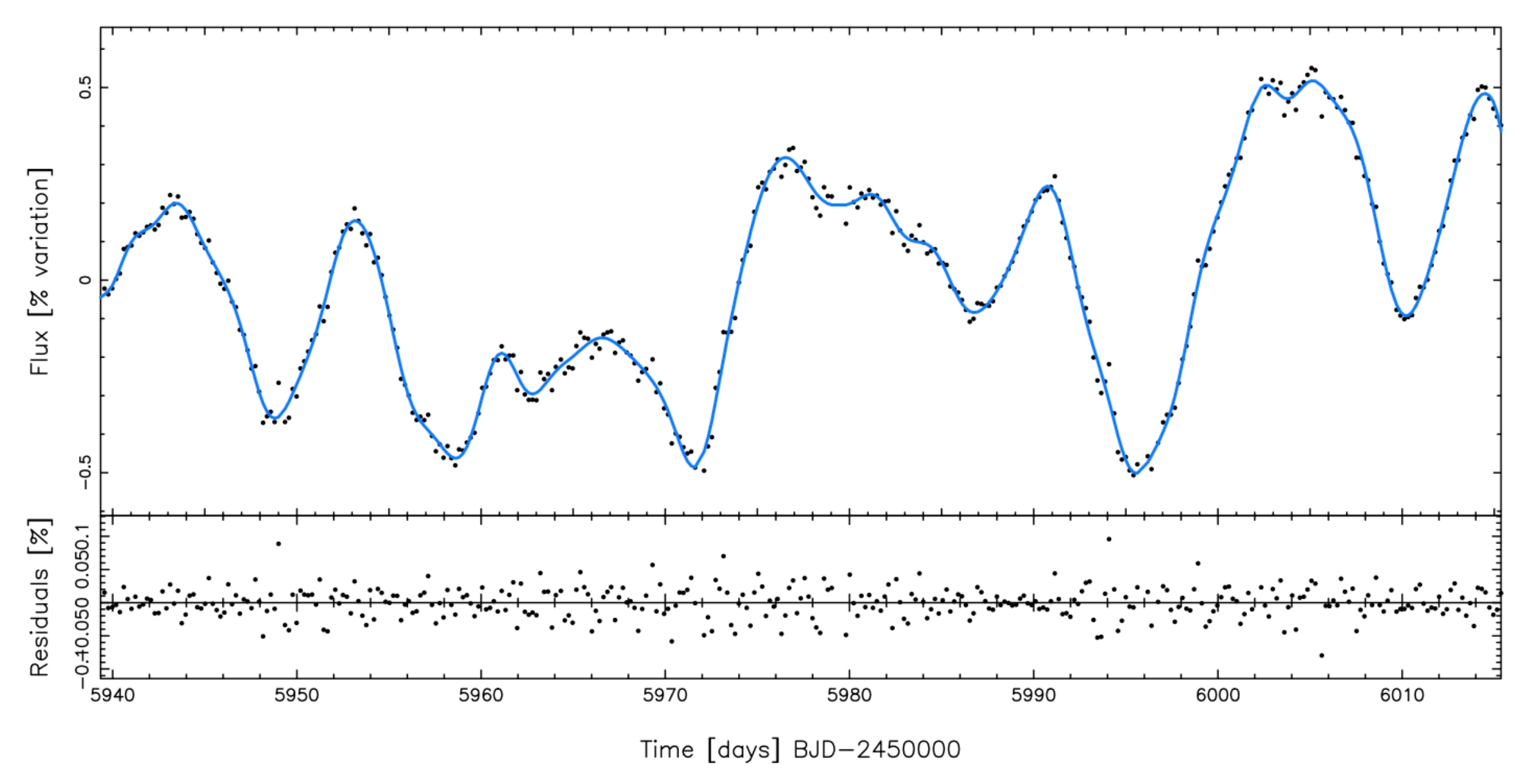}
\centering
\caption{Upper panel: CoRoT-7 light curve over the span of the 2012 RV run, with our photometric fit at each RV observation overplotted as the blue curve. Lower panel: Residuals of the fit. }
\label{lc}
\end{figure*}

\section{Observations}
\subsection{Photometry}
CoRoT-7 (average V-mag = 11.67) was observed with the CoRoT satellite \citep{Auvergne:2009en} from 2012 January 10 to March 29. Figure~\ref{lc} shows the part of the light curve which overlaps with the 2012 HARPS run. Measurements were taken in CoRoT's high cadence mode (every 32 seconds). The data were reduced with the CoRoT imagette pipeline with an optimised photometric mask in order to maximise the signal to noise of the light curve.
Further details on the data reduction are given by Barros et al. (submitted).
We binned the data in blocks of 0.07 day, which corresponds to 6176~s and is close to the orbital period of the satellite of 6184~s \citep{Auvergne:2009en} in order to average the effects of all sources of systematic errors related to the orbital motion of CoRoT.
A combined analysis of both CoRoT datasets is presented by Barros et al. (submitted). They derive the revised orbital period and epoch of first transit shown in Table~\ref{transit}. These values will be used as prior information in our MCMC simulations (see Section~\ref{priors}).

\begin{table}
  \caption{Transit information based on both CoRoT runs (preliminay results from Barros et al., submitted).}
  \begin{tabular}{@{}llrrrrlrlr@{}}
\hline
 \hline
 Period & $0.85359165\pm5.6 \times10^{-7}$ day \\
 Transit ephemeris & $2454398.07694\pm6.7 \times 10^{-4}$HJD \\
\hline
\label{transit}
\end{tabular}
\end{table}

\subsection{Spectroscopy}
\begin{figure}
\centering
\includegraphics[width = 0.5\textwidth]{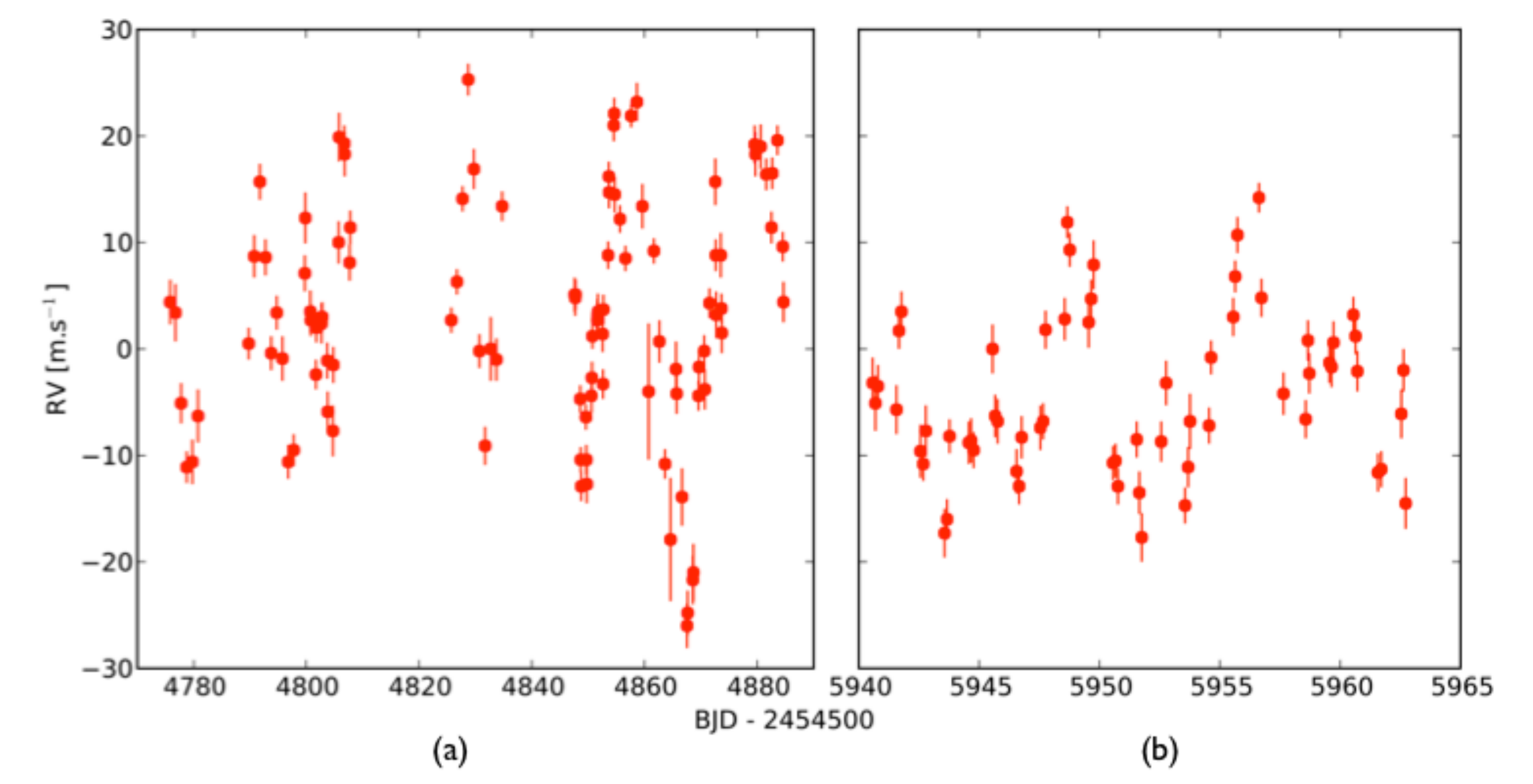}
\centering
\caption{RV variations of CoRoT-7 measured with HARPS, during the (a) $\sim$ 4 months 2008-2009 run -- (b) $\sim$ 1 month 2012 run. Note the difference in horizontal scale between the two panels.}
\label{twosets}
\end{figure}

The CoRoT-7 system was observed with the HARPS instrument \citep{Mayor:2003wva} on the ESO 3.6\,m telescope at La Silla for 26 consecutive clear nights from 2012 January 12 to February 6, with multiple well-separated measurements on each night. The 2012 RV data were reprocessed in the same way as the 2008-2009 data \citep{Queloz:2009bo} using the HARPS data analysis pipeline. The cross-correlation was performed using a K5 spectral mask.The data are given in Table~\ref{specdata}.
The median, minimum and maximum signal to noise ratio of the HARPS spectra at central wavelength 556.50 nm are 44.8, 33.8 and 56.2, respectively.
Figure~\ref{twosets} shows the RV variations of CoRoT-7 during the two campaigns. The RV variations during the second run have a smaller amplitude, implying that the star has become less active than it was in 2008-2009.


\section{A model accounting for stellar activity}\label{totalmodel}
We model the RVs as the sum of two separate contributions: one from stellar activity, and one from one or more planets. The activity component makes use of the information contained within the light curve, and is described in Sections~\ref{aigrain} to~\ref{total_activity}, while the planet(s) are assumed to follow keplerian orbits. The overall model, described in Section~\ref{mywork}, contains 3 free parameters controlling the amplitude of the different activity terms as well as 5 free parameters per planet. The manner in which we explore this parameter space and compare models with different numbers of planets is described in Section~\ref{analysis}.

\subsection{The \emph{FF'} method}\label{aigrain}
\citet{Aigrain:2012} found that RV variations induced by starspots are well reproduced by a model consisting of the product of the photometric flux $F$ and its first time derivative $F'$. It is assumed that the spots are small and limb-darkening is ignored. Spots influence the stellar RV  by suppressing the photospheric surface 
brightness at the local rotational Doppler shift of the spot. Also, in areas of high magnetic field such as faculae, which on the Sun are associated with spot groups, the convective flow is inhibited, leading to an attenuation of the convective blueshift.
This effect is thought to be the dominant contribution to the total RV signal in the Sun (Meunier, Desort and Lagrange 2010).

As derived in \citet{Aigrain:2012}, the RV perturbation $\Delta RV_{\mathrm{rot}}(t)$ to the star's RV incurred by the presence of spots on the rotating photosphere can be expressed as follows: 
\begin{equation}
\Delta RV_{\mathrm{rot}}(t) = - \frac{\dot{\Psi}(t)}{\Psi_{0}} \, \Bigl [ 1- \frac{\Psi (t)}{\Psi_{0}} \Bigr ] \, \frac{R_{\star}}{f},
\label{eqn1}
\end{equation}
where $\Psi(t)$ is the observed stellar flux, $\Psi_{0}$ is the stellar flux for a non-spotted photosphere and $\dot{\Psi}(t)$ is the first time derivative of $\Psi(t)$. $R_{\star}$ is the stellar radius. The parameter $f$ represents the drop in flux produced by a spot at the centre of the stellar disk, and can be approximated as: 
\begin{equation} 
\label{eqn2}
f \approx \, \frac{\Psi_{0} - \Phi_{\mathrm{min}}}{\Psi_{0}},
\end{equation}
where $\Phi_{\mathrm{min}}$ is the minimum observed flux, \emph{i.e.} the stellar flux at maximum spot visibility. 

The effect of the suppression of convective blueshift on the star's RV produced by starspots and magnetized areas surrounding them is given by:
\begin{equation} 
\label{eqn3}
\Delta RV_{\mathrm{conv}}(t) = \Bigl [ 1- \frac{\Psi (t)}{\Psi_{0}} \Bigr ]^{2} \, \frac{\delta V_{\rm{c}} \,\kappa}{f},
\end{equation}
where $\delta V_{\rm{c}}$ is the difference between the convective blueshift in the unspotted photosphere and that within the magnetized area, and $\kappa$ is the ratio of this area to the spot surface \citep{Aigrain:2012}. 

\subsection{Evaluating the \emph{FF'} activity basis functions}\label{modelactivity}
The flux at the time of each RV point has to be interpolated from the CoRoT light curve. In order to do this we used a Gaussian process \citep{Rasmussen:2006vz, Gibson:2011en}. A Gaussian process (GP) is a non-parametric way to model $n$ data points. Its kernel is an $n$ x $n$ covariance matrix {\bf K} in which each element contains information about how much each pair of data are correlated with each other. The matrix is determined by a model covariance function $k\,(t,t')$ whose form reflects the quasi-periodic nature of the CoRoT light curve, as evolving active regions come in and out of view:
\begin{equation}\label{cov1}
k\,(t, t') = \eta_{1}^{2} \, . \, \exp \Bigl (- \frac{(t - t')^{2}}{2\eta_{2}^{2}}  - \frac{2 \sin^{2} (\frac{\pi (t - t')}{ \eta_{3}})}{\eta_{4}^{2}}  \Bigr ).
\end{equation}
The terms $\eta_1$ (amplitude of the GP),  $\eta_2$ (timescale for growth and decay of active regions), $\eta_3$ (recurrence timescale) and $\eta_4$ (smoothing parameter) are the hyperparameters of $k(t,t')$.
The recurrence timescale was set as the rotation period of the star, which we determined by computing the discrete autocorrelation function of the light curve \citep{1988ApJ...333..646E}. We found $P_{\mathrm{rot}}\,=\,23.81 \pm 0.03$ days, which is consistent with the estimate of \citet{Leger:2009cb} of about 23 days. 
The remaining three hyperparameters were estimated through a Monte Carlo Markov Chain (MCMC), training the GP by maximising the likelihood $\mathcal{L}$ of the GP fit to the CoRoT photometry. For a dataset $\underline{y}$ \citep{Rasmussen:2006vz}:
 \begin{equation}\label{logl}
\log \mathcal{L} = - \frac{n}{2} \log(2\pi) - \frac{1}{2} \log (\vert {\bf K} + \sigma_i^2\, {\bf I}\vert) - \frac{1}{2} \, \underline{y}^{T}.\, ({\bf K}+ \sigma_i^2 \,{\bf I})^{-1}. \,\underline{y},
\end{equation}
where $\vert {\bf K} \vert$ is the determinant of the covariance matrix and acts to penalise complex models. The first term is a normalisation constant and the third term represents the $\chi^2$ of the fit. We include an additional white noise component through the term $\sigma_i^{2}\, {\bf I}$, where $\sigma_i$ is the error on each data point $y_i$ (see Table~\ref{specdata}) and {\bf I} is the identity matrix.

The best value for the hyperparameter $\eta_2$, which corresponds to the timescale for growth and decay of active regions is $\eta_2 = 20.6 \pm 2.5$ days, implying that the active regions on the stellar surface evolve on timescales similar to the stellar rotation period.
The fit is shown in the top panel of Figure~\ref{lc}. The residuals of the fit shown in the bottom panel show no correlated noise and have an rms scatter of $0.02\,\%$.
Once the covariance matrix has been calculated by training the GP on the dataset, we can use this to interpolate the value of the stellar flux and of its first time derivative at the time of each RV data point, in order to calculate $\Delta RV_{\mathrm{rot}}(t)$ and $\Delta RV_{\mathrm{conv}}(t)$.


\subsection{An additional activity basis function}
The \emph{FF'} method is likely to provide an incomplete representation of activity-induced RV variations. For example, it does not consider the broad-band photometric effect of faculae that are not physically associated with starspots; \citet{Aigrain:2012} assume that their effect on $\Delta RV_{\mathrm{rot}}$ is quite small as they tend to have low photometric contrast. Indeed, according to \citet{Lockwood:2007ir}, faculae become less important (relative to spots) in stars more active than the Sun. Faculae do, however, have a significant impact on the suppression of convective blueshift \citep{Meunier:2010hc}; indeed we find that this effect dominates the total RV contribution induced by stellar activity (see Section~\ref{activity}). There are other phenomena that the \emph{FF'} method does not account for, such as $\sim$ 50 ms$^{-1}$ inflows towards active regions recently found on the Sun (Gizon, Duvall and Larsen 2001; Gizon, Birch and Spruit 2010). Such photospheric velocity fields may affect the RV curve even if they have no detectable photometric signature. In addition, some longitudinal spot distributions have almost no photometric signature. They can nonetheless be incorporated in the RV model via a separate, flexible activity term that is not directly derived from the light curve.

We account for potential low-frequency signals not modelled by the \emph{FF'} terms by introducing an extra activity basis function that takes the form of a GP. This new GP, described by Equation~\ref{cov} below, represents an additional activity-driven RV signal, which we implicitly assume will have the same quasi-periodic covariance properties as the light curve. This GP is therefore governed by the following covariance function, with a set of hyperparameters $\theta$:
\begin{equation}\label{cov}
k\,(t, t') = \theta_{1}^{2} \, . \, \exp \Bigl (- \frac{(t - t')^{2}}{2\theta_{2}^{2}}  - \frac{2 \sin^{2} (\frac{\pi (t - t')}{ \theta_{3}})}{\theta_{4}^{2}}  \Bigr ).
\end{equation}
The amplitude of the GP, $\theta_1$ is a free parameter in our total RV model, as will be discussed in Section~\ref{mywork}. The other hyperparameters, $\theta_2$, $\theta_3$ and $\theta_4$ are equal to $\eta_2$, $\eta_3$ and $\eta_4$, respectively. This equality arises from the assumption that the frequency structure of the covariance function representing the stellar activity should be the same for both the light curve and the RV curve. Please note that for the remainder of the paper, all references to a GP refer to that described by Equation~\ref{cov} unless otherwise specified.

\subsection{Activity model}\label{total_activity}
The total RV perturbation $\Delta RV_{\mathrm{activity}}$ induced by stellar activity is then:
\begin{equation}
\Delta RV_{\mathrm{activity}} = A \Delta RV_{\mathrm{rot}} + B \Delta RV_{\mathrm{conv}} + \Delta RV_{\mathrm{additional}},
\end{equation}
where A and B are scaling factors, and the amplitude of $\Delta RV_{\mathrm{additional}}$ is controlled by the hyperparameter $\theta_1$ of Equation~\ref{cov}. In the present analysis, $R_{\star}$ (which is needed to calculate $\Delta RV_{\mathrm{rot}}$) is set to the value determined by Barros et al. (submitted). 
The values of $\delta V_{c}$ and $\kappa$ (needed for $\Delta RV_{\mathrm{conv}}$) are not known in the case of CoRoT-7 so they will be absorbed into the scaling constant $B$.


\subsection{Total RV model}\label{mywork}
Our final model consists of the three basis functions for the stellar activity as well as a keplerian signal for each one of $n_{pl}$ planets:

\begin{equation}\label{model}
\begin {aligned}
\Delta RV_{\mathrm{tot}}(t_{i})  =  RV_{0} +  \Delta RV_{\mathrm{activity}}(t_{i} , A, B, \Psi_{0}, \theta_1) \\
  + \sum \limits_{k=1}^{n_{pl}} K_{k} \bigl [\cos(\nu_{k}(t_{i}, t_{peri_k}, P_k)+\omega_{k}) + e_{k}  \cos(\omega_{k}) \bigr ] , \\
 \end{aligned}
\end{equation}
where $RV_{0}$ is a constant offset. 
The period of the orbit of planet \emph{k} is given by $P_{k}$, and its semi-amplitude is $K_{k}$. $\nu_{k}(t_{i}, t_{peri_k})$ is the true anomaly of planet \emph{k} at time $t_{i}$, and $t_{peri_k}$ is the time of periastron. Because it is difficult to constrain the argument of periastron for planets in low-eccentricity orbits, we introduce two parameters $C_{k}$ and $S_{k}$  \citep{Ford:2006ej}. They are related to the eccentricity $e_{k}$ of the planet's orbit and the argument of periastron $\omega_{k}$ as follows:
\begin{equation}
C_{k} = \sqrt{e_{k}}\, . \cos(\omega_{k}),
\end{equation}
\begin{equation}
S_{k} = \sqrt{e_{k}}\, . \sin(\omega_{k}).
\end{equation}
The use of the square root imposes a uniform prior on $e_k$, reducing the bias towards high eccentricities typically seen when defining $C_{k}$ and $S_{k}$ as $e_{k}\,\cos(\omega_{k})$ and $e_{k}\,\sin(\omega_{k})$.

The eccentricity is defined as:
\begin{equation}
e_{k} = S_{k}^{2} + C_{k}^{2},
\end{equation}
and the argument of periastron is:
\begin{equation}
\omega_{k} = \tan^{-1} (S_{k}/ C_{k}).
\end{equation}

\section{Analysis}\label{analysis}

\subsection{Periodogram analysis}\label{method1}

\begin{figure*}
\centering
\includegraphics[width = \textwidth]{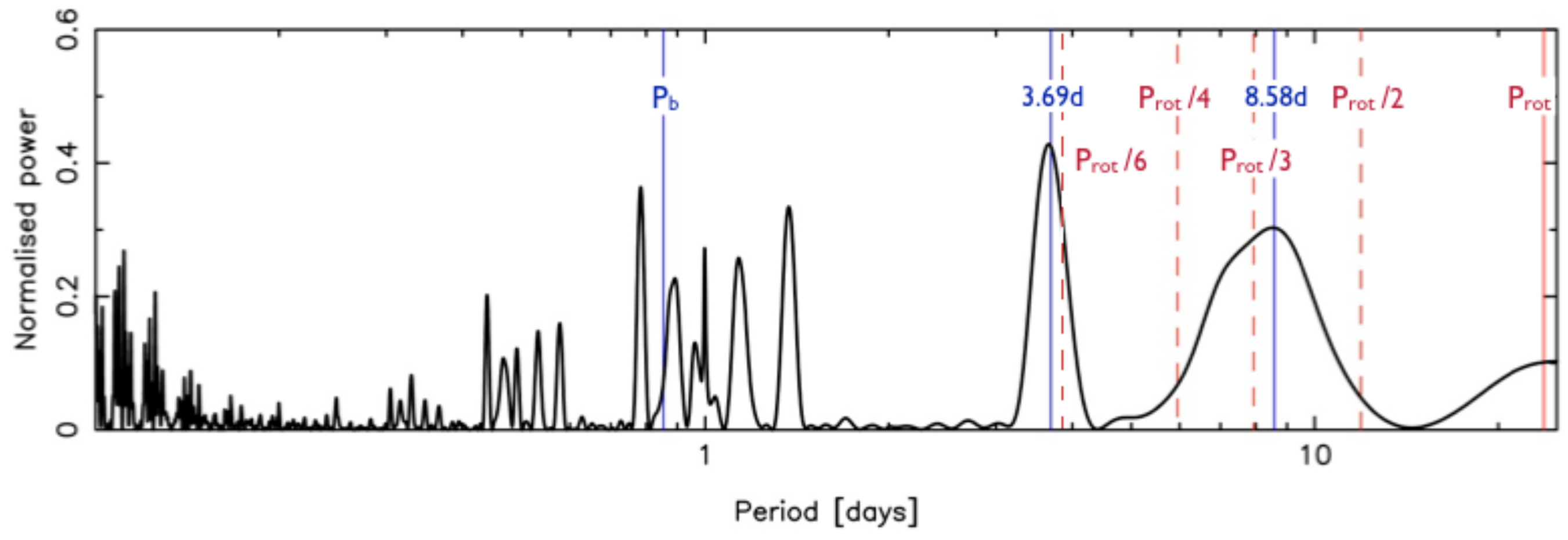}
\centering
\caption{Generalised Lomb-Scargle periodogram of the 2012 RV dataset. The stellar rotation fundamental, $P_{\rm{rot}}$, and harmonics are represented with solid and dashed lines, respectively. Also shown are the orbital period of CoRoT-7b derived from the transit analysis of Barros et al. (submitted), $P_{\rm b}$, and the periods of the two strong peaks at 3.69 and 8.58 days.}
\label{gls}
\end{figure*}

We produced a generalised Lomb-Scargle periodogram \citep{Zechmeister:2009ii} of the 2012 RV data, shown in Figure~\ref{gls}. The stellar rotation period and its harmonics are marked by the red lines (solid and dashed, respectively). 
Because the orbital period of CoRoT-7b is close to 1 day, its peak in the periodogram is hidden amongst the aliases produced by the two strong peaks at 3.69 and 8.58 days.
The peak at 3.69 days matches the period for CoRoT-7c of \citet{Queloz:2009bo}. 
We see another strong peak at a period of 8.58 days, which is close to the period found by Lanza (in prep.) of 8.29 days for the candidate planet signal CoRoT-7d, and about half a day shorter than that determined by Hatzes (in prep.) based on the same dataset. The periodogram shows that this peak is very broad and spans the whole 8-9 days range. Several stellar rotation harmonics are also present within this range, so at this stage it is not possible to conclude on the nature of this signal (this is discussed further in Section~\ref{3pl}).

\subsection{MCMC parameter fitting analysis}\label{allmcmc}
The strongest periodic signals identified in the periodogram analysis of Section~\ref{method1} were used as a starting point for an MCMC simulation (although we found that the choice of starting points does not affect the outcome of the chains). This time, the orbit of CoRoT-7b was fitted as well.

\subsubsection{Choice of priors}
The priors adopted for each parameter are given in Table~\ref{priors}.
The knee of the modified Jeffreys prior (Gregory 2007) for the semi-amplitudes of the planets was chosen to be the mean estimated error of the RV observations ($\sigma_{RV}$). Such a prior acts as a uniform prior when $K << \sigma_{RV}$, and as a Jeffreys prior for $K >> \sigma_{RV}$. This ensures that the semi-amplitudes do not get overestimated in the case of a non-detection. 
We adopt the same modified Jeffreys prior for the amplitudes $A$ and $B$ of the \emph{FF'} basis functions and the amplitude of the GP ($\theta_1$). $\theta_1$ is naturally constrained to remain low through the calculation of $\mathcal{L}$.
The orbital eccentricity of the innermost planet was constrained so that the planet's orbit remains above the stellar surface, while we imposed a simple dynamical stability criterion on the outer planets by ensuring their eccentricities were such that the orbit of each planet does not cross that of its inner neighbour.
We note that the epochs of inferior conjunction of the outer non-transiting planets (corresponding to mid-transit for a 90$^\circ$ orbit) were constrained to occur close to the inverse variance-weighted mean date of the HARPS observations in order to ensure orthogonality with the orbital periods.

\begin{table}
\caption{Prior probability densities and ranges of the parameters modelled in the MCMC procedure. The knee of the modified Jeffreys prior is given in brackets. In the case of a Gaussian distribution, the terms within brackets represent the mean $\bar{x}$ and standard deviation $\sigma$. The terms within square brackets stand for the lower and upper limit of the specified distribution; if no interval is given, no limits were placed.}
\begin{tabular}{@{}llrrrrlrlr@{}}
\hline
\hline
Parameter & Prior  \\
\hline
\rule{0pt}{0ex} \\
$RV_0$					&	Uniform 									\\
$\theta_1$				&	Modified Jeffreys ($\sigma_{RV}$) 				\\
$A$						&	Modified Jeffreys ($\sigma_{RV}$) 				\\
$B$						&	Modified Jeffreys ($\sigma_{RV}$)				\\
$\Psi_0$					&	Uniform [$\Psi_{max}$, no upper limit]			\\
$P_{\rm b}$					&	Gaussian	($0.85359165, 5.6.10^{-7}$)			\\
$t_{\rm 0b}$					&	Gaussian	($2454398.07694, 6.7.10^{-4}$)		\\
$P_{k \neq {\rm b}}$				&	 Jeffreys 									\\
$t_{0 k \neq {\rm b}}$			&	Uniform									\\
$K_k$					&	Modified Jeffreys ($\sigma_{RV}$) 				\\
$e_{\rm b}$					& 	Square root [$0, 1-\frac{R_{\star}}{a_b}$]			\\
$e_{k \neq {\rm b}}$				& 	Square root [$0, 1-\frac{a_{k-1}}{a_k} (1+e_{k-1})$]	\\
$\omega_k$				& 	Uniform [$0, 2\pi$]							\\
	
\hline
\label{priors}
\end{tabular}
\end{table}


\subsubsection{Procedure}

At every step of the chain, parameters A, B, $\Psi_{0}$, $\theta_1$, $RV_{0}$, and the orbital elements of all planets are allowed to take a random jump in parameter space. The unspotted flux level $\Psi_{0}$ is not allowed to take values less than the maximum observed flux. The two activity functions $\Delta RV_{\mathrm{rot}}$ and $\Delta RV_{\mathrm{conv}}$ are computed for every new value of $\Psi_{0}$. 
These two activity basis functions, together with the planet RVs and $RV_{0}$ are then subtracted from the data and the GP is fitted to the RV residuals. The hyperparameters $\theta_2$, $\theta_3$ and $\theta_4$ are kept fixed as they are better constrained by the light curve than the RVs, and computing them at each step of the MCMC would be cumbersome.
The likelihood $\mathcal{L}$ of the RV residuals is calculated at each step (according to Equation~\ref{logl}) in order to decide whether this set leads to a better fit than the previous set. The step is then accepted or rejected, the decision being made via the Metropolis-Hastings algorithm \citep{Metropolis:1953in}. It allows some steps to be accepted when they yield a slightly poorer fit, in order to prevent the chain from becoming trapped in a local $\mathcal{L}$ maximum and instead explore the full parameter landscape.
Once the burn-in phase is complete, \emph{i.e.} $\mathcal{L}$ becomes smaller than the median of all previous $\mathcal{L}$ \citep{Knutson:2008gl}, the chain goes through 100,000 steps, over which the standard deviations of all the parameters are calculated. These define the jump lengths in each parameter for all subsequent transition proposals.
The code goes through another 200,000 steps in order to explore the parameter landscape in the vicinity of the maximum of $\mathcal{L}$. This last phase provides the joint posterior probability distribution of all parameters of the model. The good convergence of the code was checked using the Gelman-Rubin criterion \citep{Gelman:2004tc,Ford:2006ej}, which must be smaller than 1.1 to ensure that the chain has reached a stationary state.

\section{Results and discussion}\label{discuss}
\subsection{Justification for the use of a Gaussian process}
\begin{figure}
\centering
\includegraphics[width =0.5\textwidth]{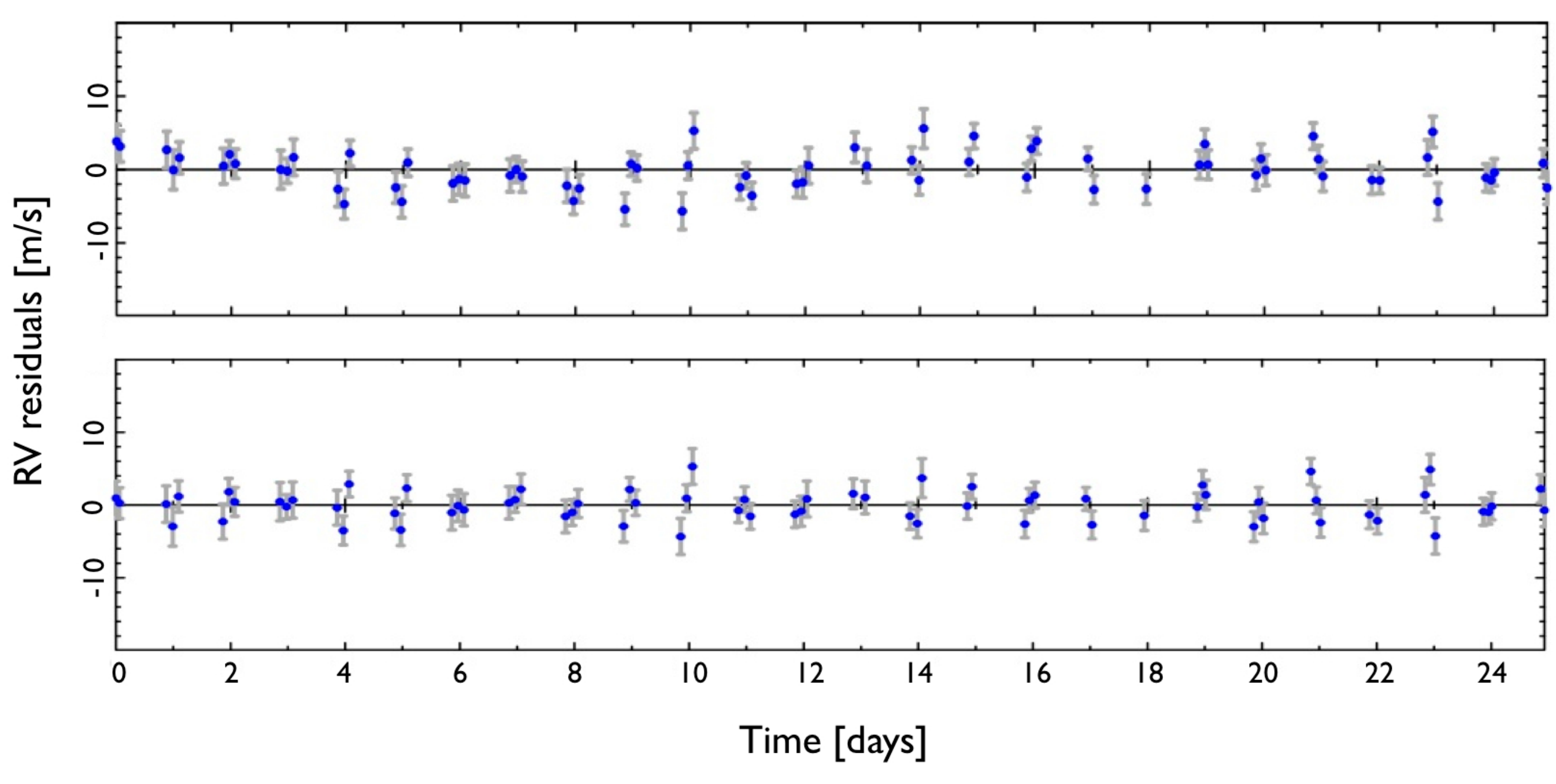}
\centering
\caption{\emph{Top:} RV residuals remaining after fitting a 3-planet + \emph{FF'} activity functions model. They contain quasi-periodic variations, and show the need to use a red noise "absorber" such as a Gaussian process. \emph{Bottom:} RV residuals after including a GP with a quasi-periodic covariance function in our RV model. The rms of the residuals, now uncorrelated, is $1.96$ m.s$^{-1}$ which is at the level of the error bars of the data.}
\label{noGP}
\end{figure}

We found that an RV model including a Gaussian process with a quasi-periodic covariance structure was the only model that would yield uncorrelated, flat residuals. Regardless of the number of planets modelled, without the inclusion of this GP the residuals always display correlated behaviour. Figure~\ref{noGP} shows the residuals remaining after fitting the orbits of CoRoT-7b, CoRoT-7c and a third keplerian, and the two basis functions of the \emph{FF'} model. We see that even the addition of a third keplerian does not absorb these variations, which appear quasi-periodic. Also, we note that a Gaussian process with a less complex, square exponential covariance function does not fully account for correlated residuals in either a 2- or 3-planet model. A comparison between a model with 2 planet orbits, the \emph{FF'} basis functions and a GP that has square exponential or quasi-periodic covariance properties yields a Bayes factor of $3.10^6$ in favour of the latter.
This implies that the active regions on the stellar surface do remain, in part, from one rotation to the next. 

\subsection{Bayesian model selection}\label{select}
We ran MCMC simulations for models with 0 (activity only), 1, and 2 planets. 
The marginal likelihood of each model is estimated from the MCMC samples using the method of \citet{Chib:2001vv} (see Appendix~\ref{app}), and is listed in the second to last row of Table~\ref{bigtable}. We also tested a 3 planet model, which is discussed in Section~\ref{3pl}.

The 2-planet model is preferred over the activity-only and 1-planet model (see the first three columns in Table~\ref{bigtable}). It is also found that a 2-planet model with free orbital eccentricities is preferred over a model with forced circular orbits by a Bayes' factor of $5.10^3$. The model with forced circular orbits is penalised mostly because of the non-zero eccentricity of CoRoT-7c. Indeed, keeping $e_b$ fixed to zero while letting $e_c$ free yields a Bayes' factor of $270$ (over a model with both orbits circular), while the Bayes' factor between models with $e_b$ fixed or free ($e_c$ free in both cases) is only 36. 
A model with no planets, consisting solely of the \emph{FF'} basis functions and a quasi-periodic GP (Model 0) is severely penalised; this attests that models with the covariance properties of the stellar activity do not absorb the signals of planets b and c.

\subsection{9-day signal: CoRoT-7d or stellar activity?}\label{3pl}
We investigated the outputs of 3-planet models in order to look for the 9-day signal present in the 2009 RV data \citep{Queloz:2009bo,Hatzes:2010jq}, whose origin has been strongly debated (cf. Section~\ref{intro} and references therein). 



First, we fitted a model comprising three keplerians, the \emph{FF'} basis functions and a Gaussian process with a quasi-periodic covariance function. We recover the two inner planets but do not detect another signal with any significance. The residuals are uncorrelated and at the level of the error bars.
We then constrained the orbital period of the third planet with a Gaussian prior centred around the period recently reported by Tuomi et al. (2014) at $P_d = 8.8999 \pm 0.0082$ days, and imposed a Gaussian prior centred at $2455949.97 \pm 0.44$ BJD on the predicted time of transit (which corresponds to the phase we determined based on the orbital period of Tuomi et al. (2014)). We recover a signal which corresponds to a planet mass of $13 \pm 5$ M$_\oplus$ and is in agreement with the mass proposed by Tuomi et al. (2014). However, the marginal likelihood of this model is $-192.5 \pm 0.7$; this is lower than the marginal likehood of the 2-planet model (Model 2, log$\mathcal{L}_{ML}$ = $-190.1 \pm 0.7$), which suggests that the addition of an extra keplerian at 9 days is not justified in view of the improvement to the fit.

Since this orbital period is very close to the second harmonic of the stellar rotation, it is plausible that the Gaussian process could be absorbing some or all of the signal produced by a planet's orbit at this period. In order to test whether this is the case, we took the residuals of Model 2 and injected a synthetic sinusoid with the orbital parameters of planet d reported by Tuomi et al. (2014). We fitted this fake dataset with a model consisting of a Gaussian process (with the same quasi-periodic covariance function as before), a keplerian and a constant offset. We find that the planet signal is completely absorbed by the keplerian model, within uncertainties -- the amplitude injected was $5.16 \pm 1.84$ m.s$^{-1}$, while that recovered is $4.97 \pm 0.35$ m.s$^{-1}$. This experiment attests that the likelihood of the model (see Equation~\ref{logl}) acts to keep the amplitude of the Gaussian process as small as possible, in order to compensate for its high degree of flexibility, and allow other parts of the model to fit the data if they are less complex than the GP. We can thus conclude that if there were a completely coherent signal close to 9 days, it would be left out by the Gaussian process and be absorbed by the third keplerian of the 3-planet model.


This signal therefore cannot be fully coherent over the span of the observations. Indeed, we see in the periodogram of the RV data in Figure~\ref{gls} that the peak at this period is broad. We note that despite the lower activity levels of the star in the 2012 datset, the 9-day period is less well determined in this dataset than in the 2008-2009 one. This peak is also broader than we would expect for a fully coherent signal at a period close to 9 days with the observational sampling of the 2012 dataset. This is likely to be caused by variations in the phase and amplitude of the signal over the span of the 2012 data.


Based on the 2012 RV dataset, we do not have enough evidence to confirm the presence of CoRoT-7d as its orbital period of 9 days is very close to the second harmonic of the stellar rotation.
Furthermore, the period measured for the 2009 dataset by \citet{Hatzes:2010jq} $P_d = 9.021 \pm 0.019$ days is not precise enough to allow us to determine whether the signals from the two seasons are in phase, as was done in the case of $\alpha$ Centauri Bb by \citet{Dumusque:2012db}.
The cycle count of orbits elapsed between the two datasets is:
$n = 1160/9.021 = 128.6$ orbits. The uncertainty is $n\,\sigma_{P_d}/P_d = n\,(0.019/9.021) = 0.27$ orbits.
Although this 1-sigma uncertainty is less than one orbit, it is big enough to make it 
impossible to test whether the signal is still coherent.
The most likely explanation, given the existing data, is that the 8-9 day signal seen in the periodogram of Figure~\ref{gls} is a harmonic of the stellar rotation.

\subsection{Best RV model: 2 planets + stellar activity}
\begin{figure*}
\centering
\includegraphics[width = 0.9\textwidth]{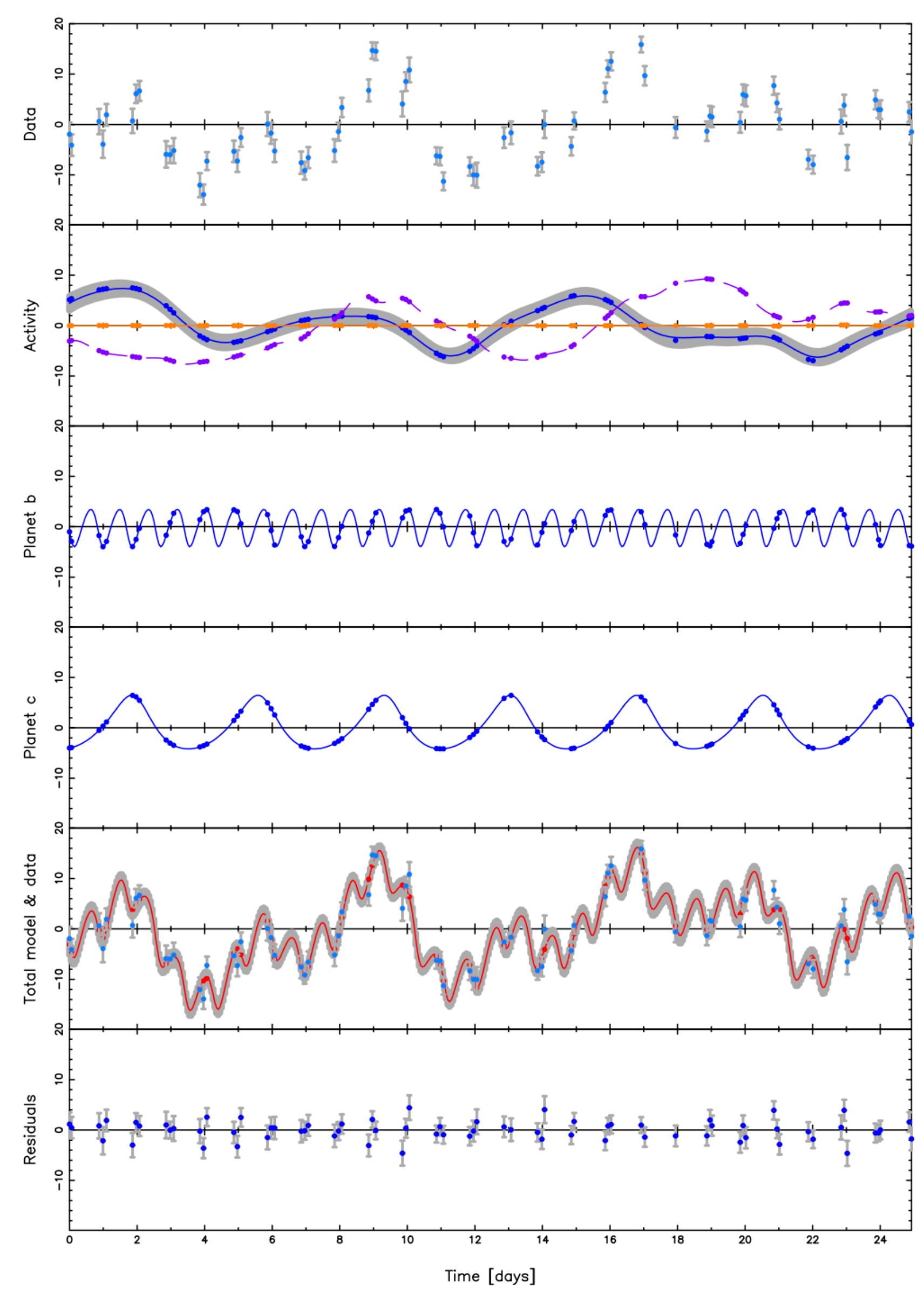}
\centering
\caption{Time series of the various parts of the total RV model for Model 2, after subtracting the star's systemic velocity $RV_0$. All RVs are in ms$^{-1}$. Panel (b): $\Delta RV_{\mathrm{rot}}$ (orange full line), $\Delta RV_{\mathrm{conv}}$ (purple dashed line) and $\Delta RV_{\mathrm{additional}}$ (blue full line with grey error band). Panel (e): the total model (red), which is the sum of activity and planet RVs, is overlaid on top of the data (blue points). Subtracting the model from the data yields the residuals plotted in panel (f).}
\label{all}
\end{figure*}

Figure~\ref{all} shows each component of the total RV model plotted over the duration of the HARPS RV campaign. We see that the suppression of convective blueshift by active regions surrounding starspots has a much greater impact on RV than flux blocked by starspots; this is discussed further in Section~\ref{activity}.


\begin{figure*}
\centering
\includegraphics[width =0.9\textwidth]{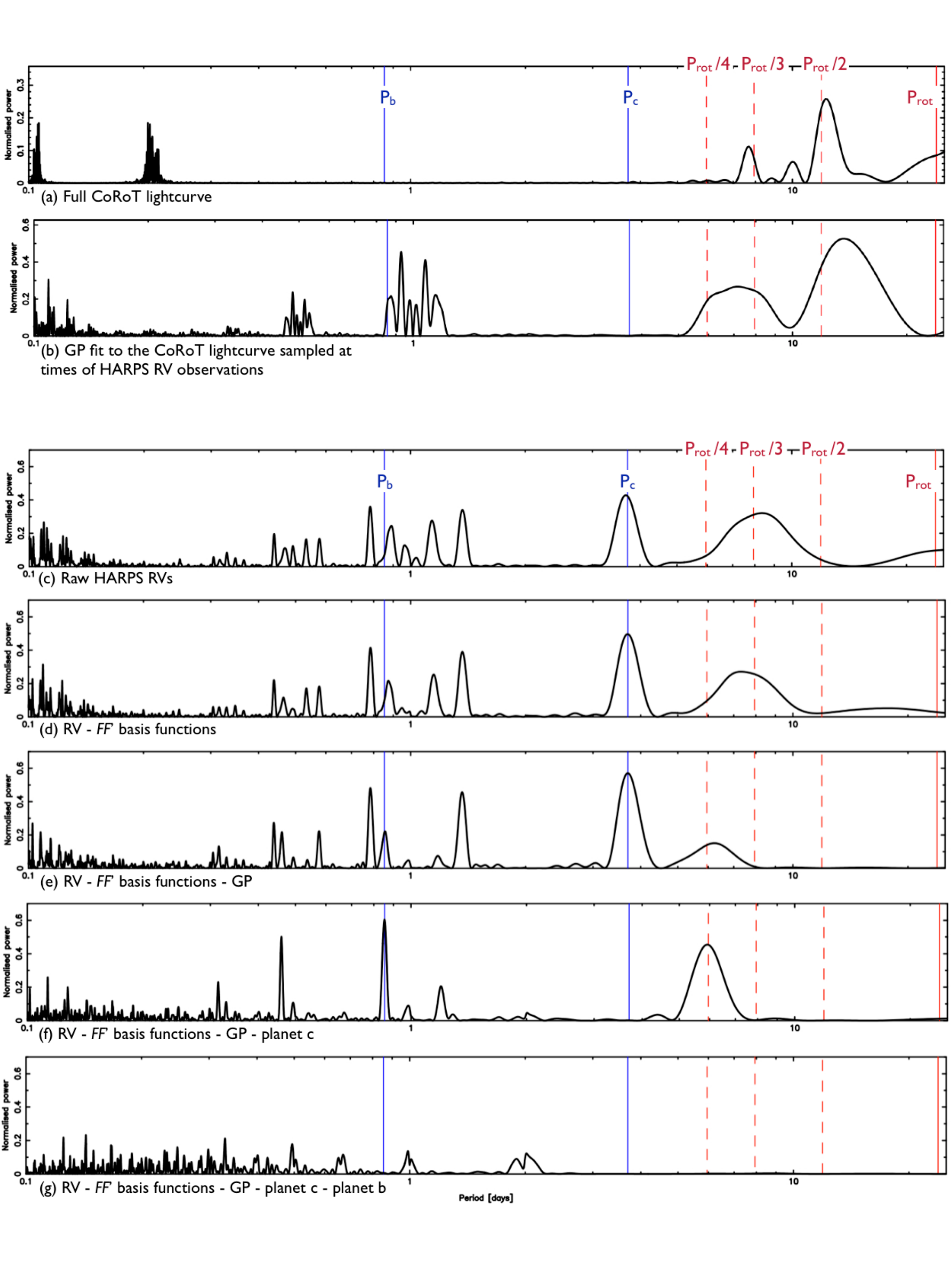}
\centering
\caption{Lomb-Scargle periodograms of: (a) the full 2012 CoRoT light curve; (b) the GP fit to the 2012 CoRoT light curve sampled at the times of RV observations; (c) the raw 2012 HARPS RV observations; (d) the RV data, from which the \emph{FF'} basis functions have been subtracted; (e) same as (d), with the Gaussian process also removed; (f) same as (e), with the signal of planet c removed; (g) same as (f), with planet b removed.}
\label{peri}
\end{figure*}

Figure~\ref{peri} shows Lomb-Scargle periodograms of the CoRoT 2012 light curve and the HARPS 2012 RV data.
Panel (a) shows the periodogram of the full CoRoT 2012 light curve, while panel (b) represents the periodogram of the GP fit to the light curve sampled at the times of the HARPS 2012 RV observations. Both periodograms reveal a stronger peak at $P_{\rm{rot}}/2$ than at $P_{\rm{rot}}$, which indicates the presence of two major active regions on opposite hemispheres of the star. This is in agreement with the variations in the light curve in Figure~\ref{lc}.
Given that suppression of convective blueshift appears to be the dominant signal, we would expect a similar frequency structure to be present in the periodogram of the RV curve (panel (c)). Indeed, we see that the stellar rotation harmonics bracket the 6 to 10 day peak in the periodogram, which has significantly greater power than the fundamental 23-day rotation signal. 
In panel (d), we remove the two \emph{FF'} basis functions. We then subtract the GP (panel (e)). We see that the GP absorbs most of the power present in the 6-10 day range. In panel (f), we have also subtracted the orbit of planet c. This removes the peaks at $P_{\rm c}$ and its 1-day alias at $\sim 1.37$ days. The peak due to CoRoT-7b now stands out along with its 1-day alias at $P = 1/(1-1/P_b) \sim 5.82$ days and harmonics $P_b/2$ and $P_b/3$. 
Finally we subtract the orbit of planet b, and are left with the periodogram of the residuals. We see that no strong signals remain except at the 1- and 2-day aliases arising from the window function of the ground-based HARPS observations.

\begin{landscape}
\begin{figure}
\centering
\includegraphics[width =1.3\textwidth]{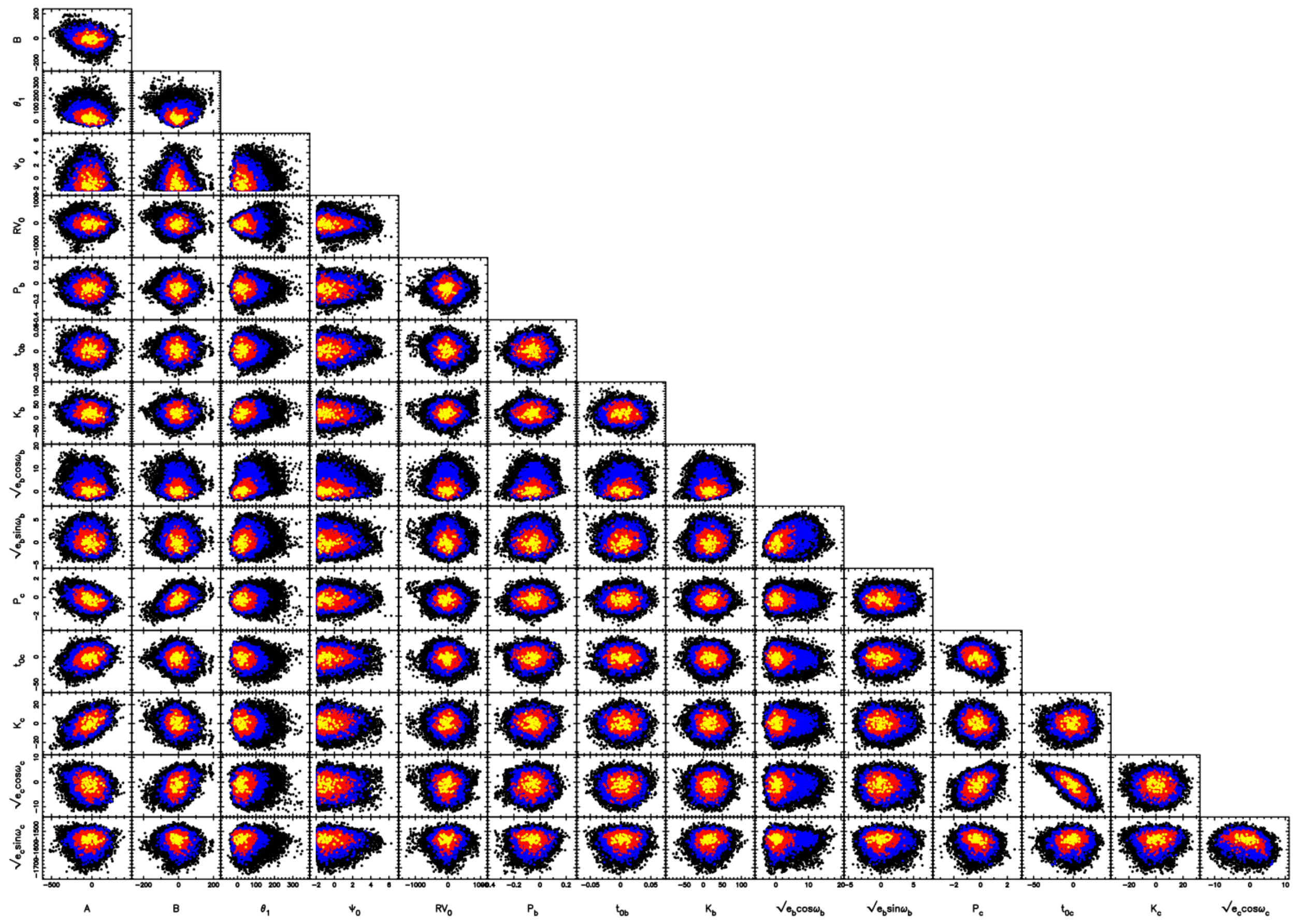}
\centering
\caption{Phase plots for the MCMC chain of Model 2 for all parameters $A$, $B$, $\theta_1$, $\Psi_0$, $RV_0$, $P_b$, $t_{0b}$, $K_b$, $\sqrt e_b \sin \omega_b$, $\sqrt e_b \cos \omega_b$, $P_c$, $t_{0c}$, $K_c$, $\sqrt e_c \sin \omega_c$, and $\sqrt e_c \cos \omega_c$. Points in yellow, red and blue are within the 1, 2 and 3-$\sigma$ confidence regions, respectively. The scale of each axis corresponds to the departure of each parameter from its value at maximum likelihood. All are expressed in percent except for $P_{\rm b}$, $t_{\rm{0b}}$ and $t_{\rm{0c}}$ which are expressed as one part per million. The distributions of $\Psi_0$ display a sharp cutoff at its minimum allowed value, which corresponds to the maximum observed flux value $\Psi_{\rm{max}}$.}
\label{shark}
\end{figure}
\end{landscape}

The posterior joint probability distributions of each pair of parameters for the 2-planet (free eccentricities) model are shown in Figure~\ref{shark}. There are no strong correlations between any of the parameters.
The $K$ amplitudes of planets b and c are found to be unaffected by the number of planets, choice of eccentric or circular orbits, or choice of activity model (all, some or none of $\Delta RV_{\mathrm{activity}}$), even when we leave $P_c$ unconstrained.
The residuals, with an rms scatter of $1.96\,$ m.s$^{-1}$ are at the level of the error bars of the data (see Table~\ref{specdata}) and show no correlated behaviour, as seen in the bottom panel of Figure~\ref{noGP}. The masses of planets b and c are presented in Sections~\ref{planetb} and~\ref{planetc}. 

\begin{table*}
\caption{Outcome of a selection of models: Model 0: stellar activity only, modelled by the \emph{FF'} basis functions and a GP with a quasi-periodic covariance function; Model 1: activity and 1 planet; Model 2: activity and 2 planets; Model 2b: activity and 2 planets with eccentricities fixed to 0.The numbers in brackets represent the uncertainty in the last digit of the value. Also given are the maximum likelihood ($\log \mathcal{L}_{\rm{max}}$), the posterior ordinate ($\hat{\pi}$) and the marginal likelihood ($\log \mathcal{L}_{\rm{ML}}$) for each model. In the last row, each model is compared to Model 2 using Bayes' factor. }
\flushleft
\begin{centering}

\begin{tabular}{llllllllll}


\hline
\hline
\rule{0pt}{4ex} & Model 0  & Model 1 & Model 2 & Model 2b \\ 

\hline
\multicolumn{3}{l}{\rule{0pt}{4ex}Stellar activity} \\
\rule{0pt}{0ex} \\

$A$ [ms$^{-1}$] 				& $-0.36 \pm 0.20$
							& $-0.35 \pm 0.21$
							& $0.06 \pm 0.13$
							& $0.06 \pm 0.12$\\

$B$ [ms$^{-1}$] 				& $0.84 \pm 1.07$
							& $-0.35 \pm 1.30$
							& $0.64 \pm 0.28$
							& $0.49 \pm 0.35$\\

$\Psi_0 / \Psi_{\mathrm{max}}$		& $1.014\pm 0.013$
							& $1.014\pm 0.012$
							& $1.014\pm 0.012$
							& $1.014\pm 0.013$ \\

$\theta_1$ [ms$^{-1}$] 			& $75 \pm 19$
							& $86 \pm 20$ 
							& $7 \pm 2$ 
							& $8 \pm 2$ \\

%
%
%
%
%

\hline
\multicolumn{1}{l}{\rule{0pt}{4ex}Planet b} \\
\rule{0pt}{0ex} \\

$P$ [days] 		& 																					& $0.85359165 (6)$ 																		& $0.85359165 (5)$ 
				& $0.85359163 (6)$ \\

$t_{0}$ [BJD - 2450000] 	& 																					& $4398.0769 (7)$ 
					& $4398.0769 (8)$
					& $4398.0769 (8)$\\

$t_{peri}$ [BJD - 2450000] & 																					& $4398.10 (7)$ 
					& $4398.21 (9)$ 
					& $4398.863 (1)$ \\

K [ms$^{-1}$] 		& 																					& $3.95 \pm 0.71$
				& $3.42 \pm 0.66$
				& $3.10 \pm 0.68$ \\

e 				& 																					& $0.17 \pm 0.09$  
				& $0.12 \pm 0.07$
				& $0$ (fixed) \\

$\omega$ [$^\circ$]  &  																					& $105 \pm 61 $ 
				& $160 \pm 140$
				& $0$ (fixed)\\

m [M$_{\oplus}$]	& 																					& $5.37 \pm 1.02$ 
				& $4.73 \pm 0.95$
				& $4.45 \pm 0.98$\\

$\rho$ [g.cm$^{-3}$] &  																					& $7.51 \pm 1.43$ 
				& $6.61 \pm 1.33$ 
				& $6.21 \pm 1.37$ \\

a [AU] 			&  																					&  $0.017 (1)$
				&  $0.017 (1)$ 
				&  $0.017 (1)$ \\

\hline
\multicolumn{1}{l}{\rule{0pt}{4ex}Planet c} \\
\rule{0pt}{0ex} \\

$P$ [days] 		& 
				&																					& $3.70 \pm 0.02$ 																		& $3.68 \pm 0.02$\\

$t_{0}$ [BJD - 2450000] 		& 					
				&																					& $5953.54 (7)$ 
				& $5953.59 (5)$ \\

$t_{peri}$ [BJD - 2450000] 	& 
				
				&
				& $5953.3 (3)$ 
				& $5952.67 (6)$ \\

K [ms$^{-1}$] 		& 
				&																					& $6.01 \pm 0.47$ 
				& $5.95 \pm 0.48$ \\

e 				& 	
				&																				 	& $0.12 \pm 0.06$ 
				& $0$ (fixed) \\

m [M$_{\oplus}$]	& 
				&																					& $13.56 \pm 1.08$
				& $13.65 \pm 1.10$ \\

a [AU] 			&  
				&																					&  $0.045 (1)$ 
				&  $0.045 (2)$ \\

%
%
%
%
%
%
%
%
%
%
%
%
%
%
%
%
%
%
			

\hline
\rule{0pt}{0ex} \\
$\log \mathcal{L}_{\rm{max}}$ 
& $-237.6 \pm 0.3$ 
& $-223.6 \pm0.5$  
& $-188.0 \pm 0.2$ 
& $ -196.28 \pm 0.04$\\

\hline
\rule{0pt}{0ex} \\
$\hat{\pi} $
& $0 \pm 1$ 
& $2 \pm 1$ 
& $2 \pm 1$ 
& $2.2 \pm 0.8$ \\

\hline
\rule{0pt}{0ex} \\
$\log \mathcal{L}_{\rm{ML}}$ 
& $-237 \pm 1$ 
& $-225 \pm 1$ 
& $-190.1 \pm 0.7$ 
& $-198.5 \pm 0.8$  \\

\hline
\rule{0pt}{0ex} \\
Bayes' factor: $B_{k,2}$
 & $4 \times 10^{-21}$ 
 & $6 \times 10^{-16}$ 
 & -
 & $2 \times 10^{-4}$  \\

\hline
\label{bigtable}
\end{tabular}

\end{centering}
\end{table*}



\subsection{CoRoT-7b}\label{planetb}

The orbital parameters of CoRoT-7b are listed in the third column of Table~\ref{bigtable}. The orbital eccentricity of $0.12 \pm 0.07$ is detected with a low significance and is compatible with the 
 transit parameters determined by Barros et al. (submitted).

As mentioned in Section~\ref{select}, the mass of CoRoT-7b is not affected by the choice of model, which attests to the robustness of this result. 
Our mass of $4.73 \pm 0.95$ M$_\oplus$ is compatible, within uncertainties, with the results found by \citet{Queloz:2009bo}, \citet{Boisse:2011bw} and \citet{Tuomi:2014tm}. It is within 2-sigma of the masses found by \citet{Pont:2010io}, \citet{Hatzes:2011fxa} and \citet{FerrazMello:2011gt}.



Using the radius found by \citet{Bruntt:2010iq}, CoRoT-7b is found to be slightly denser than the Earth ($\rho_{\oplus}$~=~5.52~g.cm$^{-3}$), with $\rho_{\rm{b}}~=~6.61 \pm 1.72$ ~g.cm$^{-3}$ (see Table~\ref{bigtable}). The reader should refer to Barros et al. (submitted) for a more detailed discussion of the density of CoRoT-7b.


\begin{figure}
\centering
\includegraphics[width = 0.5\textwidth]{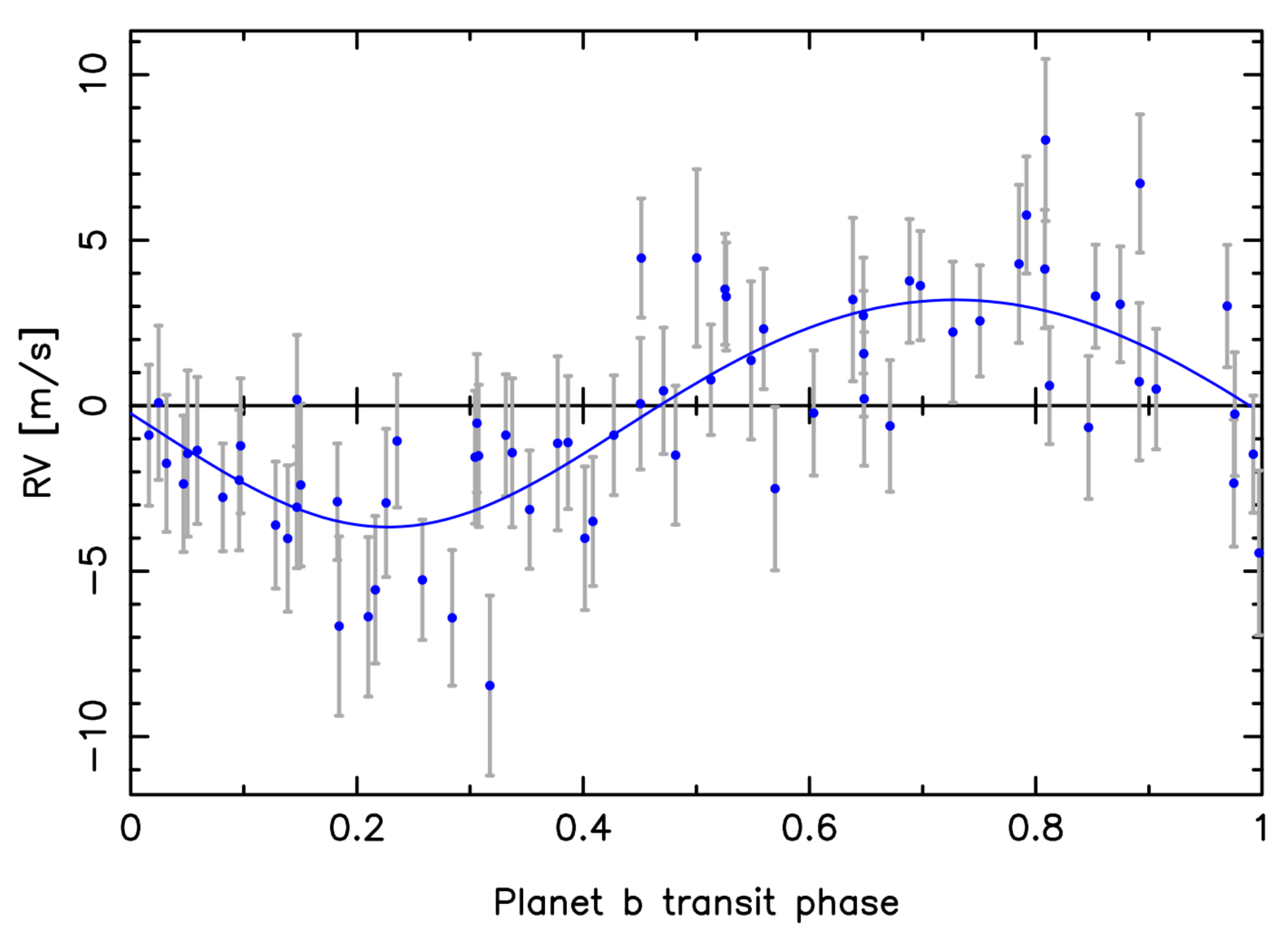}
\centering
\caption{Phase plot of the orbit of planet b for Model 2, with the contribution of the activity and planet c subtracted.}
\label{before}
\end{figure}

\subsection{CoRoT-7c}\label{planetc}
We make a robust detection of CoRoT-7c at an orbital period of 3.70 $\pm$ 0.02 days, which is in agreement with previous works that considered planet c. We estimate its mass to be 13.56$\pm$ 1.08 M$_{\oplus}$ (see Table~\ref{bigtable}). Our mass is in agreement with that given by \citet{Boisse:2011bw} and \citet{FerrazMello:2011gt}. It is just over 2-sigma lower than the mass found by \citet{Hatzes:2010jq}, and over 3-sigma greater than the mass calculated by \citet{Queloz:2009bo}.
It suggests that the harmonic filtering technique employed by \citet{Queloz:2009bo} suppresses the amplitude of the signal at this period. This may be due to the fact that $P_{\rm{c}}$ is close to the fifth harmonic of the stellar rotation, $P_{\mathrm{rot}}$/6 $\sim$ 3.9 days (see Figure~\ref{gls}), but \citet{Queloz:2009bo} only model RV variations using the first two harmonics, thus leaving $P_{\rm{c}}$ and $P_{\mathrm{rot}}$/6 entangled. \citet{FerrazMello:2011gt}, who performed a similar analysis to that of \citet{Queloz:2009bo}, mention that the proximity of $P_{\rm{c}}$ to $P_{\mathrm{rot}}$/6 may lead to underestimating the RV amplitude of CoRoT-7c by up to 0.5 ms$^{-1}$ due to beating between these two frequencies. 

We estimated the minimum orbital inclination this planet has to have in order to be transiting. Its radius $R_{\rm{c}}$ can be approximated using the formula given by \citet{Lissauer:2011fj}:
\begin{equation}
R_{\rm{c}} = \Bigl (\frac{M_{\rm{c}}}{M_{\oplus}} \Bigr )^{1/2.06} R_{\oplus},
\label{rad_eqn}
\end{equation}
where M$_{\oplus}$ and $R_{\oplus}$ are the mass and radius of the Earth.  
Using the mass for CoRoT-7c given in the third column of Table~\ref{bigtable}, we find R$_{c}$~=~3.54 R$_{\oplus}$. With this radius, CoRoT-7c would have to have a minimum orbital inclination of 83.7$^{\circ}$ in order to be passing in front of the stellar disk with respect to the observer.

CoRoT-7b's orbital axis is inclined at 79.0$^{\circ}$ to the line of sight (preliminary result of Barros et al., submitted). According to \citet{Lissauer:2011fj}, over 85\% of observed compact planetary systems containing transiting super-Earths and Neptunes are coplanar within 3$^{\circ}$. We conclude that planet c is not very likely to transit. Indeed, no transits of this planet are detected in any of the CoRoT runs. 
We infer that any planets further out from the star with a similar radius or smaller are even less likely to transit.


\begin{figure}
\centering
\includegraphics[width = 0.5\textwidth]{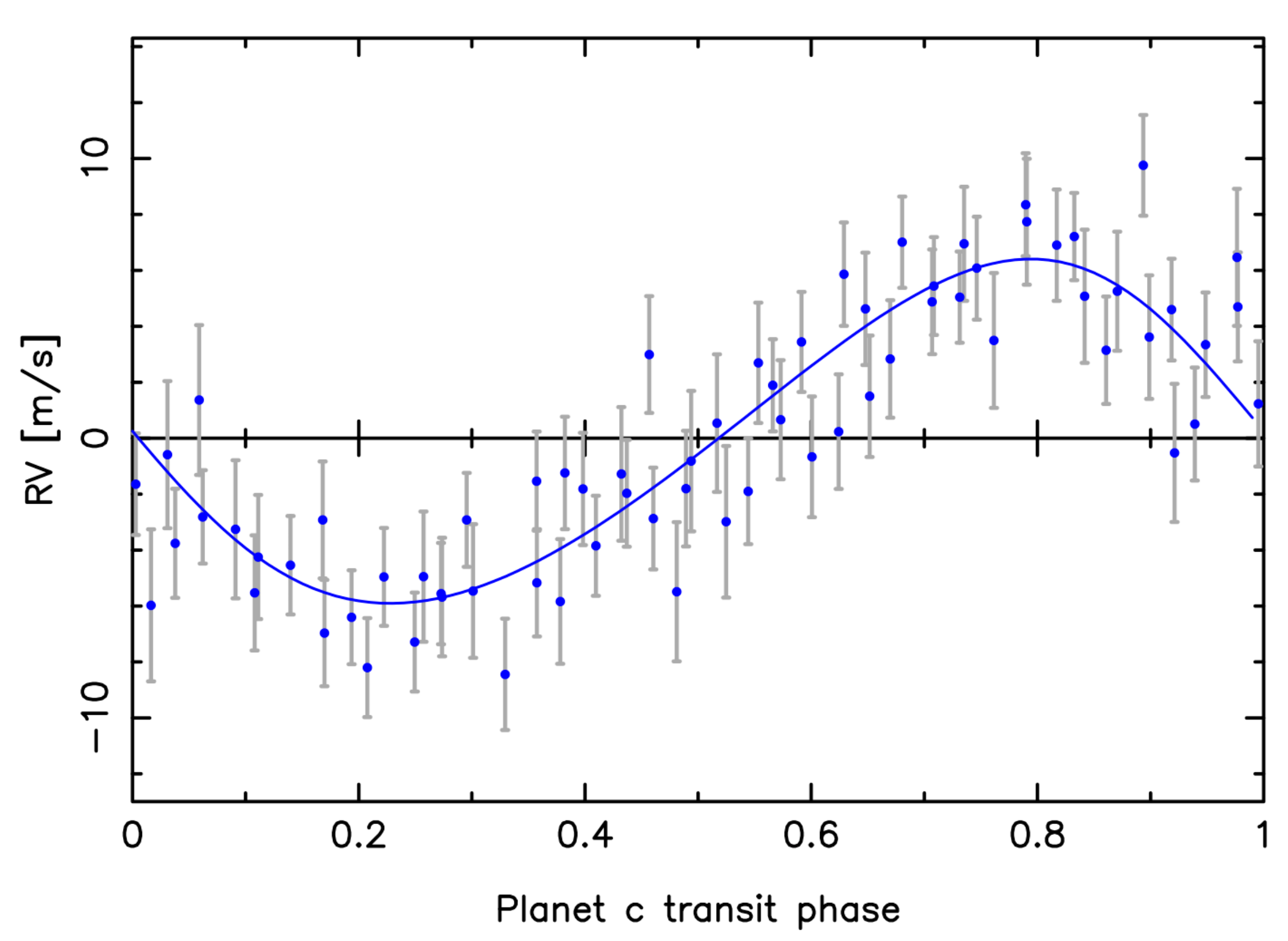}
\centering
\caption{Phase plot of the orbit of planet c for Model 2, with the contribution of the activity and planet b subtracted.}
\label{before}
\end{figure}

\subsection{The stellar activity of CoRoT-7}\label{activity}

In Model 2, the rms scatter of the total activity model is $4.86$ ms$^{-1}$ (see Figure~\ref{all}b).
For moderately active host stars such as CoRoT-7, the activity contribution largely dominates the reflex motion induced by a closely orbiting super-Earth. 

The rms scatter of $\Delta RV_{\mathrm{rot}}$ and $\Delta RV_{\mathrm{conv}}$ are $0.46$ ms$^{-1}$ and $1.82$ ms$^{-1}$, respectively.
The smaller impact of the surface brightness inhomogeneities on the RV variations could be due to the small $v \sin i$ of the star \citep{Bruntt:2010iq}, because the amplitude of these variations scales approximately with $v \sin i$ \citep{Desort:2007dt}.
This suggests that for slowly rotating stars such as CoRoT-7, the suppression of convective blueshift is the dominant contributor to the activity-modulated RV signal, rather than the rotational Doppler shift of the flux blocked by starspots.
This corroborates the findings of \citet{Meunier:2010hc} and \citet{Lagrange:2010ija}, who showed that the suppression of convective blueshift is the dominant source of activity-induced RV variations on the Sun, which is also a slowly rotating star.

We use a GP to absorb correlated residuals due to other physical phenomena occurring on timescales of order of the stellar rotation period. In the case of CoRoT-7, these combined signatures have an rms of $3.95$ ms$^{-1}$, suggesting that there are other processes than those modelled by the \emph{FF'} method at play.

\section{Conclusion}

The CoRoT-7 system was re-observed in 2012 with the CoRoT satellite and the HARPS spectrograph simultaneously.
These observations allowed us to apply the \emph{FF'} method of \citet{Aigrain:2012} to model the RV variations produced by the magnetic activity of CoRoT-7.
This approach makes use of the star's light curve and its first time derivative to model the rotational Doppler shift of the flux blocked by starspots, and the suppression of convective blueshift occurring in active regions on the stellar surface. 
If we only use the \emph{FF'} method to model the activity, we find correlated noise in the RV residuals which cannot be accounted for by a set of Keplerian planetary signals. This indicates that some activity-related noise is still present. Indeed, the \emph{FF'} method does not account for all phenomena such as the effect of limb-brightened facular emission on the cross-correlation function profile, photospheric inflows towards active regions, or faculae that are not spatially associated with starspot groups. Furthermore, some longitudinal spot distributions have almost no photometric signature. To model this low-frequency stellar signal, we use a Gaussian process with a quasi-periodic covariance function that has the same frequency structure as the light curve.

We run an MCMC simulation and use Bayesian model selection to determine the number of planets in this system and estimate their masses. We find that the transiting super-Earth CoRoT-7b has a mass of $4.73 \pm 0.95\,$M$_{\oplus}$. Using the planet radius estimated by \citet{Bruntt:2010iq}, CoRoT-7b has a density of $(6.61 \pm 1.72)(R_p/1.58$ R$_{\oplus})^{-3}$ g.cm$^{-3}$, which is compatible with a rocky composition. 
We confirm the presence of CoRoT-7c, which has a mass of $13.56 \pm 1.08\,$M$_{\oplus}$. These findings agree with the analyses made by Barros et al. (submitted), Hatzes et al. (in prep.), Lanza et al. (in prep.) and \citet{Tuomi:2014tm}.

 
We search for evidence of an additional planetary companion at a period of 9 days, as proposed by \citet{Hatzes:2010jq} following an analysis of the 2008-2009 RV dataset.
While the Lomb-Scargle periodogram of the 2012 RVs displays a strong peak in the 6-10 days range, we find that this signal is more likely to be associated with the second harmonic of the stellar rotation at $\sim 7.9$ days. 





In CoRoT-7, the RV modulation induced by stellar activity dominates the total RV signal despite the close-in orbit of (at least) one super-Earth and one sub-Neptune mass planet. 
Understanding the effects of stellar activity on RV observations is therefore crucial to improve our ability to detect low-mass planets and obtain a precise measure of their mass.

\section*{Acknowledgments}
We wish to thank the referee, S. Aigrain, for her recommendations that have greatly added to the rigour of this work. R.D. Haywood acknowledges support from an STFC postgraduate research studentship. R. Fares acknowledges support from STFC consolidated grant number ST/J001651/1. S.C.C. Barros acknowledges support by CNES, grant number 98761. A. Santerne acknowledges the support by the European Research Council/European Community under the FP7 through Starting Grant agreement number 239953. The CoRoT space mission, launched on December 27th 2006, has been developed and is operated by CNES, with the contribution of Austria, Belgium, Brazil, ESA (RSSD and Science Programme), Germany and Spain. This research has made use of NASA's Astrophysics Data System Bibliographic Services.

\bibliographystyle{mn2e} 
\bibliography{myreferences}   

\begin{thebibliography}{}

\bibitem[\protect\citeauthoryear{Aigrain, Pont \& Zucker}{Aigrain
  et~al.}{2012}]{Aigrain:2012}
Aigrain S.,  Pont F.,    Zucker S.,  2012, Monthly Notices of the Royal
  Astronomical Society, 419, 3147

\bibitem[\protect\citeauthoryear{Auvergne et~al.,}{Auvergne
  et~al.}{2009}]{Auvergne:2009en}
Auvergne M.  et~al., 2009, Astronomy and Astrophysics, 506, 411

\bibitem[\protect\citeauthoryear{Boisse, Bouchy, H{\'e}brard, Bonfils, Santos
  \& Vauclair}{Boisse et~al.}{2011}]{Boisse:2011bw}
Boisse I.,  Bouchy F.,  H{\'e}brard G.,  Bonfils X.,  Santos N.,    Vauclair
  S.,  2011, Astronomy and Astrophysics, 528, A4

\bibitem[\protect\citeauthoryear{Bruntt et~al.,}{Bruntt
  et~al.}{2010}]{Bruntt:2010iq}
Bruntt H.  et~al., 2010, Astronomy and Astrophysics, 519, A51

\bibitem[\protect\citeauthoryear{{Chib} \& {Jeliazkov}}{{Chib} \&
  {Jeliazkov}}{2001}]{Chib:2001vv}
{Chib} {Jeliazkov}, 2001, {American Statistical Association Portal :: Marginal
  Likelihood From the Metropolis--Hastings Output - Journal of the American
  Statistical Association - Volume 96, Issue 453}

\bibitem[\protect\citeauthoryear{Desort, Lagrange, Galland, Udry \&
  Mayor}{Desort et~al.}{2007}]{Desort:2007dt}
Desort M.,  Lagrange A.~M.,  Galland F.,  Udry S.,    Mayor M.,  2007,
  Astronomy and Astrophysics, 473, 983

\bibitem[\protect\citeauthoryear{Dumusque et~al.,}{Dumusque
  et~al.}{2012}]{Dumusque:2012db}
Dumusque X.  et~al., 2012, Nature, pp~--

\bibitem[\protect\citeauthoryear{Edelson \& Krolik}{Edelson \&
  Krolik}{1988}]{1988ApJ...333..646E}
Edelson R.~A.,  Krolik J.~H.,  1988, Astrophysical Journal, 333, 646

\bibitem[\protect\citeauthoryear{Ferraz-Mello, Tadeu~dos Santos, Beaug{\'e},
  Michtchenko \& Rodr{\'\i}guez}{Ferraz-Mello
  et~al.}{2011}]{FerrazMello:2011gt}
Ferraz-Mello S.,  Tadeu~dos Santos M.,  Beaug{\'e} C.,  Michtchenko T.~A.,
  Rodr{\'\i}guez A.,  2011, Astronomy and Astrophysics, 531, A161

\bibitem[\protect\citeauthoryear{Ford}{Ford}{2006}]{Ford:2006ej}
Ford E.~B.,  2006, The Astrophysical Journal, 642, 505

\bibitem[\protect\citeauthoryear{Gelman, Carlin, Stern \& Rubin}{Gelman
  et~al.}{2004}]{Gelman:2004tc}
Gelman A.,  Carlin J.~B.,  Stern H.~S.,    Rubin D.~B.,  2004, {Bayesian Data
  Analysis}.
Chapman {\&} Hall/CRC

\bibitem[\protect\citeauthoryear{Gibson, Aigrain, Roberts, Evans, Osborne \&
  Pont}{Gibson et~al.}{2011}]{Gibson:2011en}
Gibson N.~P.,  Aigrain S.,  Roberts S.,  Evans T.~M.,  Osborne M.,    Pont F.,
  2011, Monthly Notices of the Royal Astronomical Society, 419, 2683

\bibitem[\protect\citeauthoryear{Hatzes et~al.,}{Hatzes
  et~al.}{2010}]{Hatzes:2010jq}
Hatzes A.~P.  et~al., 2010, Astronomy and Astrophysics, 520, A93

\bibitem[\protect\citeauthoryear{Hatzes et~al.,}{Hatzes
  et~al.}{2011}]{Hatzes:2011fxa}
Hatzes A.~P.  et~al., 2011, The Astrophysical Journal, 743, 75

\bibitem[\protect\citeauthoryear{Hussain}{Hussain}{2002}]{Hussain:2002eg}
Hussain G.,  2002, Astronomische Nachrichten, 323, 349

\bibitem[\protect\citeauthoryear{Jeffreys}{Jeffreys}{1961}]{Jeffreys:1998tt}
Jeffreys S.~H.,  1961, {The Theory of Probability}.
Oxford University Press

\bibitem[\protect\citeauthoryear{Knutson, Charbonneau, Allen, Burrows \&
  Megeath}{Knutson et~al.}{2008}]{Knutson:2008gl}
Knutson H.~A.,  Charbonneau D.,  Allen L.~E.,  Burrows A.,    Megeath S.~T.,
  2008, The Astrophysical Journal, 673, 526

\bibitem[\protect\citeauthoryear{Kopp \& Lean}{Kopp \&
  Lean}{2011}]{Kopp:2011dv}
Kopp G.,  Lean J.~L.,  2011, Geophysical Research Letters, 38, L01706

\bibitem[\protect\citeauthoryear{Lagrange, Desort \& Meunier}{Lagrange
  et~al.}{2010}]{Lagrange:2010ija}
Lagrange A.~M.,  Desort M.,    Meunier N.,  2010, Astronomy and Astrophysics,
  512, A38

\bibitem[\protect\citeauthoryear{Lanza, Boisse, Bouchy, Bonomo \& Moutou}{Lanza
  et~al.}{2011}]{Lanza:2011bfb}
Lanza A.~F.,  Boisse I.,  Bouchy F.,  Bonomo A.~S.,    Moutou C.,  2011,
  Astronomy and Astrophysics, 533, A44

\bibitem[\protect\citeauthoryear{Lanza et~al.,}{Lanza
  et~al.}{2010}]{Lanza:2010gx}
Lanza A.~F.  et~al., 2010, Astronomy and Astrophysics, 520, A53

\bibitem[\protect\citeauthoryear{Lanza et~al.,}{Lanza
  et~al.}{2009}]{Lanza:2009gd}
Lanza A.~F.  et~al., 2009, Astronomy and Astrophysics, 493, 193

\bibitem[\protect\citeauthoryear{L{\'e}ger et~al.,}{L{\'e}ger
  et~al.}{2009}]{Leger:2009cb}
L{\'e}ger A.  et~al., 2009, Astronomy and Astrophysics, 506, 287

\bibitem[\protect\citeauthoryear{Lissauer et~al.,}{Lissauer
  et~al.}{2011}]{Lissauer:2011fj}
Lissauer J.~J.  et~al., 2011, The Astrophysical Journal Supplement Series, 197,
  8

\bibitem[\protect\citeauthoryear{Lockwood, Skiff, Henry, Henry, Radick,
  Baliunas, Donahue \& Soon}{Lockwood et~al.}{2007}]{Lockwood:2007ir}
Lockwood G.~W.,  Skiff B.~A.,  Henry G.~W.,  Henry S.,  Radick R.~R.,  Baliunas
  S.~L.,  Donahue R.~A.,    Soon W.,  2007, The Astrophysical Journal
  Supplement Series, 171, 260

\bibitem[\protect\citeauthoryear{Mayor et~al.,}{Mayor
  et~al.}{2003}]{Mayor:2003wva}
Mayor M.  et~al., 2003, The Messenger, 114, 20

\bibitem[\protect\citeauthoryear{Metropolis, Rosenbluth, Rosenbluth, Teller \&
  Teller}{Metropolis et~al.}{1953}]{Metropolis:1953in}
Metropolis N.,  Rosenbluth A.~W.,  Rosenbluth M.~N.,  Teller A.~H.,    Teller
  E.,  1953, The Journal of Chemical Physics, 21, 1087

\bibitem[\protect\citeauthoryear{Meunier, Desort \& Lagrange}{Meunier
  et~al.}{2010}]{Meunier:2010hc}
Meunier N.,  Desort M.,    Lagrange A.~M.,  2010, Astronomy and Astrophysics,
  512, A39

\bibitem[\protect\citeauthoryear{Pont, Aigrain \& Zucker}{Pont
  et~al.}{2010}]{Pont:2010io}
Pont F.,  Aigrain S.,    Zucker S.,  2010, Monthly Notices of the Royal
  Astronomical Society, 411, 1953

\bibitem[\protect\citeauthoryear{Queloz et~al.,}{Queloz
  et~al.}{2009}]{Queloz:2009bo}
Queloz D.  et~al., 2009, Astronomy and Astrophysics, 506, 303

\bibitem[\protect\citeauthoryear{Queloz et~al.,}{Queloz
  et~al.}{2001}]{Queloz:2001be}
Queloz D.  et~al., 2001, Astronomy and Astrophysics, 379, 279

\bibitem[\protect\citeauthoryear{Rasmussen \& Williams}{Rasmussen \&
  Williams}{2006}]{Rasmussen:2006vz}
Rasmussen C.~E.,  Williams C. K.~I.,  2006, {Gaussian Processes for Machine
  Learning}.
MIT Press

\bibitem[\protect\citeauthoryear{Schrijver}{Schrijver}{2002}]{Schrijver:2002ht}
Schrijver C.~J.,  2002, Astronomische Nachrichten, 323, 157

\bibitem[\protect\citeauthoryear{Tuomi, Anglada-Escud{\'e}, Jenkins \&
  Jones}{Tuomi et~al.}{2014}]{Tuomi:2014tm}
Tuomi M.,  Anglada-Escud{\'e} G.,  Jenkins J.~S.,    Jones H. R.~A.,  2014,
  eprint arXiv:1405.2016

\bibitem[\protect\citeauthoryear{Zechmeister \& K{\"u}rster}{Zechmeister \&
  K{\"u}rster}{2009}]{Zechmeister:2009ii}
Zechmeister M.,  K{\"u}rster M.,  2009, Astronomy and Astrophysics, 496, 577

\end{thebibliography}

\appendix

\section{Model selection}\label{bayesian}\label{app}
We ran MCMC chains for several different models and selected the best one using Bayesian statistics.

\subsection{Bayes' factor}
Given a dataset ${\bf y}$, consider two models $\mathcal{M}_i$ and $\mathcal{M}_j$. In order to determine which one is the simplest  but still gives the best fit to the data, one can compare the two models by estimating their posterior odds ratio:
\begin{equation}
\frac{P(\mathcal{M}_i | {\bf y})}{P(\mathcal{M}_j | {\bf y})} = \frac{Pr(\mathcal{M}_i)}{Pr(\mathcal{M}_j)} \cdot \frac{m({\bf y} | \mathcal{M}_i)}{m({\bf y} | \mathcal{M}_j)},
\end{equation}
where the first factor on the right side of the equation is the prior odds ratio. In this analysis, all models that are tested have the same prior information, so this ratio is just 1. This leaves us with the second part of the right side of the equation. It is the ratio of the marginal likelihoods $m$ of each model, and is known as Bayes' factor.

The marginal likelihood $m$ of a dataset {\bf y} given a model $\mathcal{M}_i$ with a set of parameters $\theta_i$ can be written as:
\begin{equation}
m ({\bf y} | \mathcal{M}_i) = \int f ({\bf y} | \mathcal{M}_i, \theta_i) \,  \pi_i (\theta_i | \mathcal{M}_i) \, d\theta_i,
\end{equation}
where $f ({\bf y} | \mathcal{M}_i, \theta_i)$ is the likelihood function $\mathcal{L}$. The term $\pi_i (\theta_i | \mathcal{M}_i)$ accounts for the prior distribution of the parameters and can be incorporated as a penalty to $\mathcal{L}$. 
According to \citet{Chib:2001vv}, it is possible to write:
\begin{equation}
m ({\bf y} | \mathcal{M}_i) = \frac{f ({\bf y} | \mathcal{M}_i, \theta_i) \,  \pi (\theta_i | \mathcal{M}_i)}{\pi (\theta_i | {\bf y}, \mathcal{M}_i)}.
\end{equation}
The denominator $\pi (\theta_i | {\bf y}, \mathcal{M}_i)$ is the posterior ordinate, which we estimate using the posterior distributions of the parameters resulting from MCMC chains.

\subsection{Posterior ordinate}
According to \citet{Chib:2001vv}, the posterior ordinate $\hat{\pi} (\theta_i | {\bf y})$ can be evaluated by comparing the mean transition probability for a series of $M$ jumps from any given $\theta_i$ \emph{to} a reference $\theta_*$, to the mean acceptance value for a series of $J$ transitions \emph{from} $\theta_*$.
This can be written as:
\begin{equation}\label{11}
\hat{\pi} (\theta_* | {\bf y}) = \frac{M^{-1} \sum \limits_{i=1}^M \alpha (\theta_i, \theta_* | {\bf y}) \cdot q (\theta_i, \theta_* | {\bf y})}{J^{-1} \sum \limits_{j=1}^J \alpha (\theta_*, \theta_j|{\bf y})},
\end{equation}
where $\alpha (\theta_i, \theta_*|{\bf y})$ is the acceptance probability of the chain from one  parameter set $\theta_i$ to another set $\theta_*$. The proposal density $q (\theta_i, \theta_* | {\bf y})$ from one step $\theta_i$ to another  $\theta_*$ is equal to:
\begin{equation}
q (\theta_i, \theta_* | {\bf y}) = \exp \Bigl [- \sum \limits_{k=1}^{K} \Bigl (\frac{\theta_i - \theta_*}{\sigma_{\theta_i}} \Bigr )^2 / 2 \Bigr ].
\end{equation}
The summation inside the exponential term is carried out over all $K$ parameters of the model, in other words over each parameter contained within a set $\theta$.

If we choose $\theta_*$ to be the best parameter set of the whole MCMC chain, then the acceptance probability $\alpha (\theta_i, \theta_* | {\bf y})$ is 1, and Equation~\ref{11} is much simplified.

\subsection{Marginal likelihood}
One can obtain $\mathcal{L}_{ML}$ by subtracting the posterior ordinate from the maximum likelihood value of the whole MCMC chain:
\begin{equation}
\log \mathcal{L}_{ML} = \log \mathcal{L}_{best} - \log \hat{\pi}.
\end{equation}

When the number of model parameters becomes very large, the summation on the numerator of equation A4 is dominated by a relatively small fraction of points in the Markov chain that happen to lie close to the maximum likelihood value. A large number of trials is therefore needed to arrive at a reliable estimate of $\hat{\pi}$ . We estimated the uncertainty in the posterior ordinate by running the chains several times and determining the variance empirically. These uncertainties are listed in Table~\ref{bigtable}.

Once $\mathcal{L}_{ML}$ is known we can compute Bayes' factor for a pair of models. The posterior ordinate acts to penalise models that have too many parameters.  \citet{Jeffreys:1998tt} found that the evidence in favour of a model is decisive if Bayes' factor exceeds 150, strong if it is in the range of 150-20, positive for 20-3 and not worth considering if lower than 3.

\begin{table*}
\caption{HARPS 2012 data for CoRoT-7, processed in the same way as the 2008-2009 data \citep{Queloz:2009bo}. From left to right are given: Julian date, radial-velocity $RV$, the estimated error $\sigma_{RV}$ on the RV, the full width at half-maximum ($FWHM$) and the line bisector of ($BIS$) of the cross-correlation function (as defined in \citet{Queloz:2001be}), the Ca II activity indicator log($R'_{HK}$) and its error $\sigma_{log(R'_{HK})}$.}
\begin{center}
\begin{minipage}{126mm}
\begin{tabularx}{\textwidth}{llllllllllll}
Julian Date & \textit{RV} & \textit{$\sigma_{RV}$} & \textit{FWHM} & \textit{BIS} & log($R'_{HK}$) & $\sigma_{log(R'_{HK})}$ \\
    
     [Day] BJD\_UTC & [km.s$^{-1}$] & [km.s$^{-1}$] & [km.s$^{-1}$] & [km.s$^{-1}$] &       &  \\
    \hline
     \rule{0pt}{0ex} \\
    2'455'939.69948 & 31.18031 & 0.00233 & 6.45633 & 0.01199 & -4.6990 & 0.0180 \\
    2'455'939.76024 & 31.17814 & 0.00212 & 6.46445 & 0.01966 & -4.7188 & 0.0173 \\
    2'455'940.57499 & 31.18283 & 0.00251 & 6.46592 & 0.01956 & -4.6982 & 0.0210 \\
    2'455'940.68929 & 31.17833 & 0.00271 & 6.45650 & 0.03679 & -4.7789 & 0.0283 \\
    2'455'940.79456 & 31.18415 & 0.00215 & 6.46568 & 0.01359 & -4.7204 & 0.0180 \\
    2'455'941.56490 & 31.18294 & 0.00241 & 6.45065 & 0.01822 & -4.7635 & 0.0245 \\
    2'455'941.66870 & 31.18832 & 0.00184 & 6.45273 & 0.02814 & -4.7365 & 0.0137 \\
    2'455'941.77024 & 31.18890 & 0.00199 & 6.45558 & 0.02559 & -4.7510 & 0.0169 \\
    2'455'942.56139 & 31.17631 & 0.00263 & 6.45472 & 0.01846 & -4.6730 & 0.0191 \\
    2'455'942.67696 & 31.17626 & 0.00167 & 6.45427 & 0.02055 & -4.7071 & 0.0095 \\
    2'455'942.78412 & 31.17705 & 0.00247 & 6.45564 & 0.03818 & -4.7222 & 0.0221 \\
    2'455'943.56090 & 31.17020 & 0.00239 & 6.44767 & 0.02167 & -4.7187 & 0.0215 \\
    2'455'943.66570 & 31.16834 & 0.00199 & 6.44649 & 0.02202 & -4.7482 & 0.0163 \\
    2'455'943.76867 & 31.17497 & 0.00177 & 6.45494 & 0.02110 & -4.7508 & 0.0152 \\
    2'455'944.56671 & 31.17690 & 0.00213 & 6.44517 & 0.02596 & -4.7237 & 0.0180 \\
    2'455'944.66911 & 31.17499 & 0.00216 & 6.44351 & 0.02340 & -4.7166 & 0.0167 \\
    2'455'944.77370 & 31.17966 & 0.00185 & 6.44134 & 0.02035 & -4.7206 & 0.0150 \\
    2'455'945.56098 & 31.18232 & 0.00238 & 6.45457 & 0.01149 & -4.7319 & 0.0221 \\
    2'455'945.66736 & 31.18053 & 0.00213 & 6.45674 & 0.01439 & -4.7275 & 0.0178 \\
    2'455'945.77208 & 31.17698 & 0.00221 & 6.44160 & 0.02874 & -4.7497 & 0.0213 \\
    2'455'946.55742 & 31.17466 & 0.00222 & 6.44724 & 0.00971 & -4.7694 & 0.0214 \\
    2'455'946.66311 & 31.17309 & 0.00176 & 6.45013 & 0.01661 & -4.7358 & 0.0131 \\
    2'455'946.76840 & 31.17567 & 0.00209 & 6.45653 & 0.01566 & -4.7467 & 0.0181 \\
    2'455'947.54531 & 31.17707 & 0.00223 & 6.45981 & 0.02810 & -4.7581 & 0.0210 \\
    2'455'947.66174 & 31.18084 & 0.00179 & 6.45909 & 0.01563 & -4.7334 & 0.0133 \\
    2'455'947.76281 & 31.18561 & 0.00191 & 6.46437 & 0.02390 & -4.7700 & 0.0185 \\
    2'455'948.55706 & 31.18901 & 0.00217 & 6.46402 & 0.01550 & -4.7355 & 0.0188 \\
    2'455'948.66364 & 31.19692 & 0.00163 & 6.46248 & 0.02318 & -4.7389 & 0.0114 \\
    2'455'948.76718 & 31.19676 & 0.00175 & 6.46623 & 0.02778 & -4.7548 & 0.0157 \\
    2'455'949.55411 & 31.18631 & 0.00247 & 6.46951 & 0.02427 & -4.8253 & 0.0283 \\
    2'455'949.65555 & 31.19076 & 0.00187 & 6.46200 & 0.02620 & -4.7545 & 0.0149 \\
    2'455'949.75824 & 31.19305 & 0.00245 & 6.46655 & 0.03246 & -4.7434 & 0.0243 \\
    2'455'950.56227 & 31.17601 & 0.00168 & 6.46236 & 0.03281 & -4.7404 & 0.0130 \\
    2'455'950.66816 & 31.17590 & 0.00175 & 6.45419 & 0.01683 & -4.7400 & 0.0131 \\
    2'455'950.76859 & 31.17096 & 0.00177 & 6.45464 & 0.02764 & -4.7633 & 0.0163 \\
    2'455'951.54884 & 31.17391 & 0.00182 & 6.43438 & 0.03045 & -4.7528 & 0.0158 \\
    2'455'951.65576 & 31.17223 & 0.00207 & 6.43799 & 0.02989 & -4.7971 & 0.0188 \\
    2'455'951.75704 & 31.17219 & 0.00246 & 6.44706 & 0.02311 & -4.7875 & 0.0271 \\
    2'455'952.56523 & 31.17963 & 0.00204 & 6.44030 & 0.01523 & -4.7580 & 0.0187 \\
    2'455'952.77021 & 31.18059 & 0.00225 & 6.44565 & 0.02541 & -4.7385 & 0.0225 \\
    2'455'953.55597 & 31.17395 & 0.00182 & 6.43480 & 0.01000 & -4.7119 & 0.0146 \\
    2'455'953.68468 & 31.17475 & 0.00195 & 6.45180 & 0.00298 & -4.7323 & 0.0152 \\
    2'455'953.76300 & 31.18222 & 0.00268 & 6.44616 & 0.02591 & -4.7364 & 0.0272 \\
    2'455'954.55404 & 31.17790 & 0.00181 & 6.44536 & 0.00875 & -4.7410 & 0.0154 \\
    2'455'954.63792 & 31.18295 & 0.00168 & 6.46153 & 0.01177 & -4.7168 & 0.0118 \\
    2'455'955.55847 & 31.18862 & 0.00189 & 6.46166 & 0.02000 & -4.7413 & 0.0161 \\
    2'455'955.63894 & 31.19331 & 0.00165 & 6.46554 & 0.01452 & -4.7438 & 0.0120 \\
    2'455'955.73279 & 31.19476 & 0.00179 & 6.44739 & 0.00867 & -4.7347 & 0.0148 \\
    2'455'956.62463 & 31.19811 & 0.00156 & 6.46579 & 0.02657 & -4.7147 & 0.0106 \\
    2'455'956.72897 & 31.19191 & 0.00192 & 6.47400 & 0.02364 & -4.7015 & 0.0149 \\
    2'455'957.64372 & 31.18164 & 0.00206 & 6.47184 & 0.02857 & -4.7539 & 0.0179 \\
    2'455'958.56684 & 31.18091 & 0.00192 & 6.47635 & 0.02301 & -4.7447 & 0.0163 \\
    2'455'958.65850 & 31.18392 & 0.00201 & 6.47078 & 0.02216 & -4.7459 & 0.0171 \\
    2'455'958.71729 & 31.18374 & 0.00201 & 6.47376 & 0.02909 & -4.6738 & 0.0158 \\
    2'455'959.55361 & 31.18266 & 0.00205 & 6.48019 & 0.01837 & -4.7103 & 0.0164 \\
    2'455'959.64103 & 31.18817 & 0.00201 & 6.46959 & 0.02289 & -4.7087 & 0.0149 \\
    2'455'959.72214 & 31.18793 & 0.00210 & 6.46640 & 0.02085 & -4.7060 & 0.0177 \\
    2'455'960.54986 & 31.18993 & 0.00180 & 6.48016 & 0.02823 & -4.6961 & 0.0133 \\
    2'455'960.64222 & 31.18652 & 0.00182 & 6.48139 & 0.01905 & -4.6868 & 0.0124 \\
    2'455'960.71808 & 31.18327 & 0.00202 & 6.47342 & 0.03245 & -4.6930 & 0.0161 \\
\end{tabularx}
\end{minipage}
\end{center}
\label{specdata}
\end{table*}%


\begin{table*}
\contcaption{}
\begin{center}
\begin{minipage}{126mm}
\begin{tabularx}{\textwidth}{llllllllllll}
Julian Date & \textit{RV} & \textit{$\sigma_{RV}$} & \textit{FWHM} & \textit{BIS} & log($R'_{HK}$) & $\sigma_{log(R'_{HK})}$ \\
    
     [Day] BJD\_UTC & [km.s$^{-1}$] & [km.s$^{-1}$] & [km.s$^{-1}$] & [km.s$^{-1}$] &       &  \\
    \hline
     \rule{0pt}{0ex} \\
    2'455'961.57133 & 31.17531 & 0.00190 & 6.46406 & 0.01978 & -4.7245 & 0.0148 \\
    2'455'961.71143 & 31.17428 & 0.00177 & 6.47095 & 0.01987 & -4.7172 & 0.0140 \\
    2'455'962.54203 & 31.18283 & 0.00239 & 6.48479 & 0.01274 & -4.7236 & 0.0204 \\
    2'455'962.63340 & 31.18606 & 0.00209 & 6.47948 & 0.02189 & -4.7266 & 0.0166 \\
    2'455'962.72313 & 31.17570 & 0.00249 & 6.47445 & 0.01550 & -4.7286 & 0.0235 \\
    2'455'963.55846 & 31.18712 & 0.00187 & 6.48049 & 0.01961 & -4.7051 & 0.0135 \\
    2'455'963.64853 & 31.18517 & 0.00163 & 6.47868 & 0.02397 & -4.6832 & 0.0101 \\
    2'455'963.70438 & 31.18517 & 0.00184 & 6.47691 & 0.02248 & -4.6987 & 0.0143 \\
    2'455'964.55809 & 31.18475 & 0.00195 & 6.48780 & 0.03089 & -4.7172 & 0.0148 \\
    2'455'964.62532 & 31.18077 & 0.00224 & 6.48025 & 0.02273 & -4.7404 & 0.0182 \\
    2'455'964.70360 & 31.17426 & 0.00272 & 6.47958 & 0.03171 & -4.7357 & 0.0249 \\
    
\hline
\end{tabularx}
\end{minipage}
\end{center}
\end{table*}%

\bsp

\label{lastpage}

\end{document}